\documentclass[draft,onecolumn, 10pt]{IEEEtran}
\usepackage{amsmath,amssymb,threeparttable,placeins,enumerate,caption}
\usepackage{mathtools,relsize,cite}
\usepackage[pdftex]{graphicx}
\usepackage{epstopdf}
\usepackage[usenames]{color}
\usepackage{amsfonts}
\usepackage{latexsym}
\usepackage{subfigure,multirow,makecell,stfloats}

\definecolor{burntorange}{rgb}{0.8, 0.33, 0.0}

\usepackage{algorithm}
\usepackage[noend]{algpseudocode}

\newtheorem{theorem}{{{\textit{Theorem}}}}

\newtheorem{lemma}{{{\textit{Lemma}}}}
\newtheorem{corollary}{{{{\textit{Corollary}}}}}

\newtheorem{definition}{{{\textit{Definition}}}}

\newtheorem{remark}{{{\textit{Remark}}}}

\newtheorem{example}{{{\textit{Example}}}}

\newtheorem{construction}{{{\textit{Construction}}}}
\begin{document}
	
	
	\title{New Spectrally Constrained Sequence Sets with Optimal {Periodic} Cross-Correlation}
	\author{Zhifan Ye,
       Zhengchun Zhou,
       Zilong Liu,
        Xiaohu Tang,
        and Pingzhi Fan, Fellow, IEEE.
\thanks{Z. Ye and Z. Zhou are with the School of Mathematics, and also with  the Key Lab of Information Coding and Wireless Comms., Southwest Jiaotong University, Chengdu, 610031, China. Email: yzffjnu@163.com, zzc@swjtu.edu.cn.}
\thanks{Z. Liu is with the School of Computer Science and Electronics Engineering, University of Essex, Colchester CO4 3SQ, U.K.. E-mail: zilong.liu@essex.ac.uk.}
\thanks{X. Tang and P. Fan are with the Key Lab of Information Coding and Wireless Comms., Southwest Jiaotong University, Chengdu, 611756, China. E-mail: xhutang@swjtu.edu.cn, pzfan@swjtu.edu.cn.}
}
\maketitle
	
	\begin{abstract}
		Spectrally constrained sequences (SCSs) play an important role in modern communication and radar systems operating over non-contiguous spectrum. Despite numerous research attempts over the past years, very few works are known on the constructions of optimal SCSs with low cross-correlations. In this paper, we address such a major problem by introducing  a unifying framework to construct unimodular SCS families using circular Florentine rectangles (CFRs) and interleaving techniques. By leveraging the uniform power allocation in the frequency domain for all the admissible carriers (a necessary condition for beating the existing periodic correlation lower bound of SCSs), we present a tighter correlation lower bound and show that it is achievable by our proposed SCS families including multiple SCS sets with zero correlation zone properties.
	\end{abstract}
	

\section{Introduction}
\subsection{Background}
Designing sequences with good correlation properties for various communication and radar systems has been a significant research topic for several decades. Specifically, sequences with low auto- and cross-correlations are highly desirable for a number of applications, such as active sensing, ranging, channel estimation, synchronization and spread spectrum communications, etc \cite{fanbook,golomb}. Conventional sequences are typically designed with the assumption of contiguous spectral bands. Specifically, a separate contiguous spectral band is assigned to a communication or radar application with guard bands inserted at both ends so as to avoid interference from adjacent bands. However, such a luxury is hard to continue \cite{zhao}. On one hand, modern communication and radar applications demand larger amount of spectral bands to provide higher data rates and/or enhanced sensing performances. On the other hand, the current radio spectrum (particularly the sub-6 GHz band) is becoming increasingly congested and fragmented owing to the explosive growth of wireless applications and communication/sensing devices.

For increased spectral utilization efficiency, as an instance, an overlay cognitive radio network \cite{haykin,yucek,haykin1} keeps searching for unused spectral bands to serve new users (called secondary users), whilst ensuring minimal interference to the licensed users.
When multiple secondary users are to be deployed over several non-contiguous frequency bands, SCSs with low cross-correlations are desired. Most of the known constructions of SCSs are obtained by numerical optimization. Two algorithms are developed in \cite{he2010} to construct unimodular SCSs for applications in cognitive radar. \cite{tsai2011} studied SCSs with both low autocorrelation and low peak-to-average power ratios using Gerchberg-Saxton (GS) algorithm \cite{ger1972}. \cite{rowe2014} proposed fast Fourier transform (FFT) based SHAPE algorithm to design SCSs. The waveform trade-offs between the achievable signal to interference plus noise ratio, spectral shape, and the resulting autocorrelation function were studied in \cite{aubry2014}. SCSs for cognitive code division multiple access communications were developed in \cite{hu2014} by time-frequency analysis. A remarkable progress was made in \cite{song2016} by applying majorization-minimization technique to design SCSs.

SCSs with good correlation properties are also useful in attaining random access over non-contiguous spectrum. In a cellular network, the physical random access channel (PRACH) conveys uplink synchronization signals in order to establish an initial access from a user equipment (UE) to a base station (BS) \cite{pit2020}. The network should support multiple PRACH signals, which are preamble sequences, to enable simultaneous detection of multiple UEs accessing the network. In order for the BS to correctly detect multiple preambles {and to estimate} their timing for synchronization, the preambles should have low auto- and cross-correlation properties {over certain} range of access delays in the same cell. Besides, the cross-correlations of multiple preambles from distinct cells, also known as inter-set correlations, should be as small as possible to suppress/mitigate the interference between different cells. In LTE and 5G New Radio systems, the preambles are constructed from Zadoff-Chu (ZC) sequences with different cyclic shifts and root indices. Some sequence families with similar properties have been reported in the literature (see \cite{tang2006}, \cite{popvic2010}, \cite{zhou2018} and references therein). However, these sequences are all {designed} over contiguous spectral bands.
In addition, most available sequences are contiguous {which are inapplicable} in some non-contiguous spectral scenarios, such as the interlaced transmission of 3GPP LTE enhanced licensed-assisted access (eLAA) and New Radio (NR) in unlicensed bands. In 3GPP 5G NR R16, the physical uplink control channels (PUCCH) formats can be extended to the interlaced transmission for unlicensed bands \cite{3gpp19}. To satisfy such requirement of non-contiguous spectral applications, Sahin and Yang proposed a general construction of Golay complementary pairs {(GCPs) in \cite{Sahin20} which can cater for both contiguous and non-contiguous spectrum bands. Recently, a new construction of non-contiguous complementary sequences has been developed in \cite{Shen21} for more flexible resource allocation in the frequency domain. In general, however, there is a paucity of sequences which can be used in such spectrally constrained applications.}
By contrast, this paper is mainly concerned with the multiple SCS sets with optimal inter-set cross-correlation which may be used to implement the detection and timing estimation in a spectrally constrained communication scenario.

\subsection{Motivations and Contributions}
It is noted that many well-known correlation lower bounds, such as Welch bounds \cite{welch}, Sarwate bounds \cite{sarwate}, and Tang-Fan-Matsufuji bounds \cite{tangfan}, are applicable to traditional sequences with contiguous spectral bands only. A correlation lower bound to measure the optimality of SCSs was first derived in \cite{tsai}. Such a bound was generalized and extended in \cite{liu2018} by convex optimization in the frequency domain for single- and multi-channel SCSs. For SCSs with zero power leakage over all the forbidden frequency slots (i.e., sometimes also called carriers in this paper), it is shown in \cite{liu2018} that the corresponding periodic correlation lower bound is met with equality if and only if uniform power allocation is applied to those admissible frequency slots. By leveraging this condition, we aim to go one step further to tighten that periodic correlation lower bound. The key observation behind the derivation is that uniform power allocation over all the admissible carriers yields a constant sum of correlation squares when all the typical time-shifts are counted. Such an observation allows us to differentiate the auto- and cross- correlation lower bounds of SCSs separately for a tighter lower bound (see \textit{Theorem 1} and \textit{Theorem 2}).

As far as the systematic constructions of optimal SCSs are concerned, to the best of our knowledge, \cite{tsai} pioneered the first ZCZ SCS family by allocating a comb-like sequence in the frequency domain. In 2018, \cite{liu2018} constructed SCSs with minimum autocorrelation values by judiciously choosing certain ternary frequency-domain duals with zero periodic autocorrelation sidelobes. Recently, \cite{Tian2020} proposed an analytical construction of single-channel polyphase SCS families whose maximum periodic correlation magnitude asymptotically achieves the correlation lower bound derived in \cite{liu2018}. However, the size of that sequence family is at most the minimal prime factor of the sequence period.

\begin{figure}
\centering
	\includegraphics[draft=false,width=12cm]{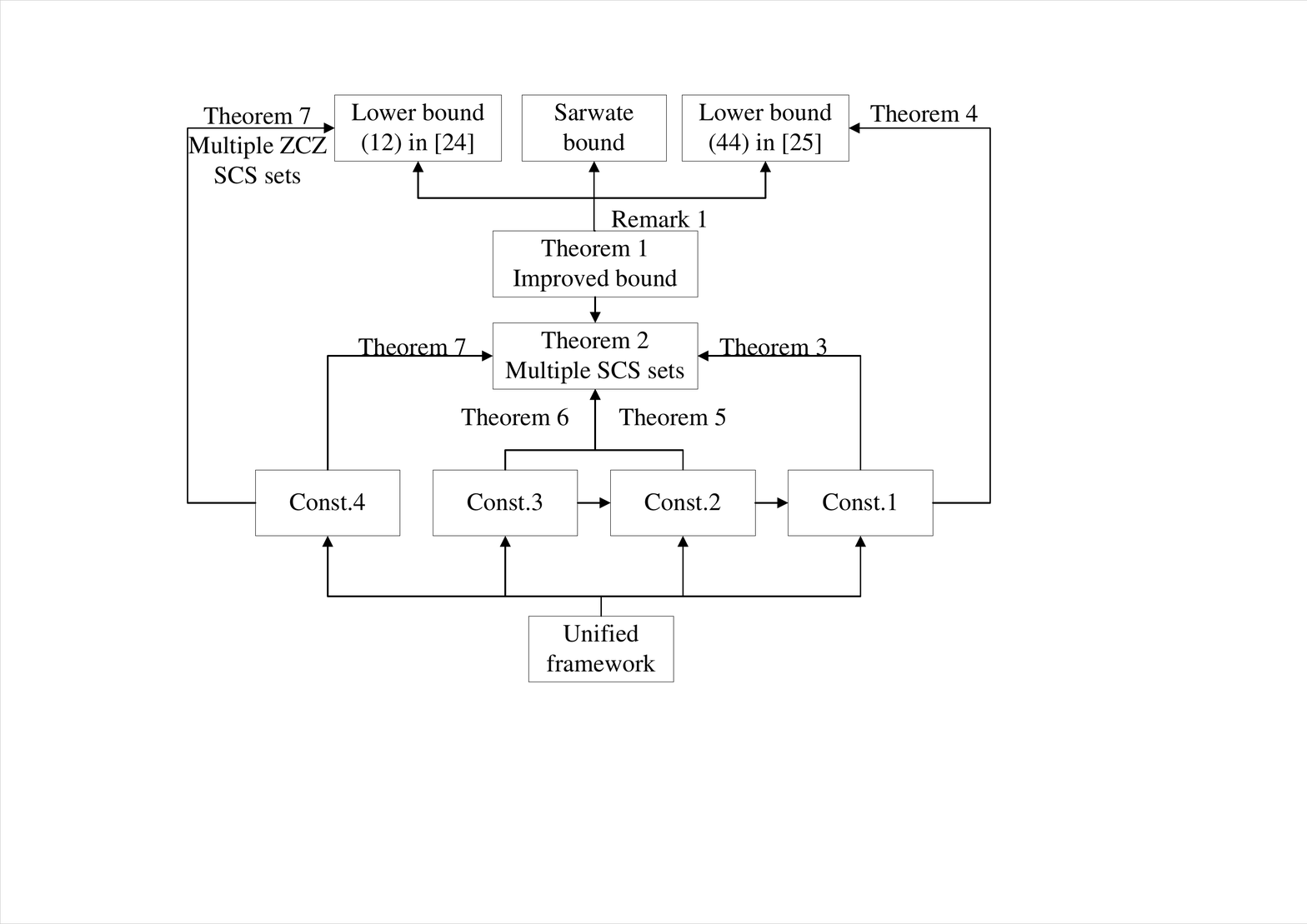}
	\caption{An illustration of the connections of the main results in this paper.}\label{figure}
\end{figure}

Motivated by \cite{liu2018} and \cite{Tian2020}, we first propose a novel construction of SCSs by applying {circular Florentine rectangles (CFRs)} in the time domain. Florentine rectangles are a combinatorial concept which has been extensively studied since 1989 \cite{golombF,taylorF,songF,thesisF}. Florentine rectangles of order $k \times n$, are a special class of matrices having $k$ rows each of which contains $n$ distinct symbols exactly once and for any pair of distinct symbols ($a,b$) and $1\leq m <n$, there is at most one row for which $b$ is $m$ steps right to $a$. A CFR is defined when circular rows are considered \cite{thesisF}. With the aid of CFRs, we show that the set size of the proposed SCS family is larger than that in \cite{Tian2020} while maintaining the minimum correlation magnitude, making it closer to the lower bound in \cite{liu2018} (see \textit{Theorem 3} and \textit{Theorem 4}). In addition, we design a generic framework through interleaving technique in the frequency domain (see \textit{Section V}). By selecting the base sequence appropriately, some known SCS families can be obtained by our framework.
Interestingly,  by applying the inverse of CFRs in the frequency domain, we obtain new SCSs with low correlation and more flexible forbidden frequency slots (see \textit{Construction 2-4}).
It is worth mentioning that traditional designs can only produce a small number of sequences because of the linear structure used. {By leveraging the combinatorial structure of CFRs, our constructions lead to large-sized SCSs for the support of more users.}
In addition, analytical construction of SCSs with flexible spectral null constraints structure and low-correlation properties is challenging due to a paucity of effective tools.
In view of this, we show that new spectral null constraints can be supported by the proposed construction  combining interleaving and cyclic difference sets (see \textit{Theorem 6} and \textit{Theorem 7}).
In particular, among these new constructions, we obtain multiple SCS sets with ZCZ properties meeting both the set size upper bound (derived based on \cite{tsai}) and our improved inter-set correlation lower bound of SCSs.
Finally, we summarize the main results of this paper in Fig. \ref{figure} in order to illustrate their inter-connections. For example, the arrow below \textit{Theorem 3} in the Fig. \ref{figure} means that \textit{Construction 1} achieves the lower bound in  \textit{Theorem 2} through \textit{Theorem 3}. The same can be said for \textit{Theorem 4}, \textit{Theorem 5}, \textit{Theorem 6} and \textit{Theorem 7}. More information is also provided in the subsequent constructions, theorems and remarks.

\subsection{Organization}
The rest of the paper is organised as follows. In Section II, we revisit some preliminary definitions, introduce the concepts of SCSs and CFRs. Also, we review several properties and  constructions of CFRs in the literature. In Section III, we revisit some known bounds of SCSs proposed in \cite{tsai} and \cite{liu2018}, and then derive an improved lower bound for minimum cross-correlation of multiple SCS sets. Section IV proposes new optimal SCS families using CFRs in the time domain. In Section V, we carry out time-frequency analysis under a newly proposed unifying framework for the design of novel SCS families. In Section VI, some optimal SCS sets are presented by our frame work, including optimal multiple ZCZ SCS sets meeting the improved lower bounds. Finally, we conclude the paper in Section VII.

\section{Preliminaries}
Before we begin, let us define the notations that we will be used throughout this paper.
\begin{itemize}
	\item  $L$ is an integer.
	\item $\mathbb{Z}_L$ denotes the ring of integers modulo $L$.
	\item {$\omega_{L}=e^{\frac{2\pi \sqrt{-1}}{L}}=e^{\frac{2\pi \textit{i}}{L}}$ is a primitive $L$-th complex root of unity.}
	\item $||.||_F$ denotes the Frobenius norm.
	\item $\langle.\rangle_L$ denotes the integer modulo $L$.
	\item $\lfloor a/b \rfloor$ denotes the largest integer not greater than $a/b$.
	\item $\mathfrak{C}$ denotes a family of sequence sets.
	\item $\mathcal{C}$ denotes a sequence set.
	\item $C$ denotes a sequence.
	\item $x^*$ denotes the complex conjugate of $x$.
	\item $\mathcal{F}_L=[f_{i,j}]_{i,j=0}^{L-1}$ is a (scaled) discrete Fourier transform (DFT) matrix of order $L$, i.e., $f_{i,j}=\frac{1}{\sqrt{L}}\omega_L^{-ij}$, for $0\leq i,j \leq L-1$.
	\item $C=[c_0,c_1,\dots,c_{L-1}]$ is a time domain sequence of length $L$, whose corresponding frequency domain dual is $\widehat{C}=[\widehat{c}_0,\widehat{c}_1,\dots,\widehat{c}_{L-1}]=C\mathcal{F}_L$.
\end{itemize}

\begin{definition}
	Let $C=\left[c_{0}, c_{1}, \cdots, c_{L-1}\right]$ and $D=\left[d_{0}, d_{1}, \cdots, d_{L-1}\right]$ be two length-$L$ complex-valued sequences. The periodic cross-correlation function (PCCF) between $C$ and $D$ is defined as
	\begin{equation}
		{\theta}_{C, D}(\tau)=\sum_{t=0}^{L-1} c_{t} d_{\langle t+\tau\rangle_{L}}^{*}=\sum_{f=0}^{L-1} \hat{c}_{f} \hat{d}_{f}^{*}\omega_L^{-f\tau}, ~ {0 \leq \tau \leq L-1}.
	\end{equation}
	When $C=D$, it is called periodic autocorrelation function which is denoted by $\theta_{C}(\tau)$.
\end{definition}

\begin{definition}[ZCZ sequence set]
	Let $\mathcal{C}$ be a family of $M$ sequences of length $L$, i.e., $\mathcal{C}=\{C_{0},C_{1},\ldots,C_{M-1}\}$, where
	$C_j=[c_{j,0},c_{j,1},\dots,c_{j,L-1}]$ denotes the $j$-th constituent sequence of length $L$. $\mathcal{C}$ is said to be an $(M,L,Z)$- ZCZ sequence set with ZCZ width $Z$, if it satisfies the following conditions:
	\begin{equation}
		\begin{split}
			{\theta}_{C_i}(\tau)&=0, \text{ for }0\leq i <M \text{ and } 0<\tau <Z;\\
			{\theta}_{C_i,C_j}(\tau)&=0, \text{ for }0\leq i\neq j <M \text{ and } 0\leq\tau <Z.
		\end{split}
	\end{equation}
\end{definition}

\subsection{{Multiple} Spectrally Null Constrained Sequences}

Let $\mathfrak{C}$ be a set of $K$ sequence sets, each consisting of $M$ sequences of length-$L$, i.e., $\mathfrak{C}=\{\mathcal{C}^0,\mathcal{C}^1,\dots,\mathcal{C}^{K-1}\}$, where $\mathcal{C}^{i}=\{C_{0}^{i},C_{1}^{i},\ldots,C_{M-1}^{i}\}$ and
$C^i_j=[c^i_{j,0},c^i_{j,1},\dots,c^i_{j,L-1}]$ denotes the $j$-th constituent sequence of the $i$-th set and each sequence of $\mathcal{C}^i$ has identical energy of $L$, i.e., $||C^i_j||^2_F=L$. Unimodular sequences are desired as they lead to improved power transmission efficiency. Let us consider a cognitive radio/radar system. The entire spectrum is assumed to be divided into $L$ carriers. Let $[d_0,d_1,\dots,d_{L-1}]$ denote a ``carrier marking vector" which displays the status of all carriers of the system. To be specific, $d_t=1$ if the $t$-th carrier is available, otherwise, $d_t=0$. Let $\Omega$ denotes the ``spectral constraint", the set of all forbidden carrier positions, i.e., $\Omega=\{t:d_t=0,t\in \mathbb{Z}_L\}$.
Formally, for any sequence $C^i_j$ in $\mathfrak{C}$, denote $\widehat{C}^i_j=[\widehat{c}^i_{j,0},\widehat{c}^i_{j,1},\dots,\widehat{c}^i_{j,L-1}]$ its frequency domain dual sequence. For minimum integrated correlation squares of SCSs, it is shown in \cite{liu2018} that uniform power allocation should be adopted to all the admissible carriers (i.e., carriers whose positions are complementary to that of the forbidden carriers). Throughout this paper, we consider such a setting for power allocation. We also assume that all the sequences in an SCS set follow the same spectral-null constraint. Therefore, we consider the SCS set $C$ which satisfies the following condition:
	\begin{equation}
		|\widehat{c}^i_{j,k}|^2=\left\{
    \begin{array}{ll}
      \frac{L}{L-|\Omega|}, & k\not\in \Omega; \\
      0, & k\in \Omega;
    \end{array}
  \right.
	\end{equation}
for any $~0\leq i \leq K-1$, $0\leq j \leq M-1$.

\subsection{Interleaving Technique}
Following the terminology in \cite{golomb}, let $\mathcal{A}$ be an $M\times N$ matrix as follows:
\begin{equation}
	\mathcal{A}=\begin{bmatrix}
		a_{0,0} & a_{0,1} & \dots & a_{0,N-1}\\
		a_{1,0} & a_{1,1} & \dots & a_{1,N-1}\\
		\vdots & \vdots & \ddots & \vdots \\
		a_{M-1,0} & a_{M-1,1} & \dots & a_{M-1,N-1}
	\end{bmatrix}_{M\times N}.
\end{equation}
 Then the interleaving sequence of $\mathcal{A}$, denoted by $A=[a_0,a_1,\dots,a_{MN-1}]$ is a length-$MN$ sequence constructed by concatenating the rows of $\mathcal{A}$. In other words,
\begin{equation}
	a_{Mi+j}=a_{i,j}, \text{ for }0\leq i <M,~ 0\leq j <N.
\end{equation}
Here, $\mathcal{A}$ is called the base matrix of $A$.

\subsection{Difference Sets}
In this subsection, we give the definition of cyclic difference set over $\mathbb{Z}_{v}$ as follows.
For any subset $\mathcal{D} = \{d_{0}, d_{1},\ldots,d_{k-1}\} \in \mathbb{Z}_{v}$, the difference function of $\mathcal{D}$ is defined as
$$
d_{\mathcal{D}}(\varepsilon) = |(\varepsilon + \mathcal{D})\cap \mathcal{D}|,~~ \varepsilon \in \mathbb{Z}_{v} .
$$
Then $\mathcal{D}$ is said to be a $(v, k, \lambda)$ cyclic difference set if and only if $d_{\mathcal{D}}(\varepsilon)$ takes on the value $\lambda$ for $v-1$ times when $\varepsilon$ ranges over the nonzero elements of $\mathbb{Z}_{v} $.

\subsection{Circular Florentine Rectangles}

In this subsection, we introduce the definition of CFRs which is available in the literature \cite{songF,thesisF}.

\begin{definition}\label{tuscandef}
	A Tuscan-$k$ rectangle of order $r \times N$ has $r$ rows and $N$ columns such that
	\begin{enumerate}
		\item[C1:] Each row is a permutation of the $N$ symbols and
		\item[C2:] For any two distinct symbols $a$ and $b$ and for each $1\leq m \leq k$, there is at most one row in which $b$ is $m$ steps to the right of $a$.
	\end{enumerate}
	When $k=N-1$, it is called a Tuscan-$(N-1)$ rectangle or Florentine rectangle. When the circularly-shifted versions of the rows of a Florentine rectangle satisfying the condition that  $b$ is $N-m$ steps to the right of $a$ is equivalent to the fact that $b$ is $m$ steps to the left of $a$, it is called a CFR.
Recently, Zhang and Helleseth defined the CFRs by using matrix notation in \cite{Zhang}. Each row, denoted by $\pi_{i}$ for $0 \leq i \leq r-1$, is a permutation of $\mathbb{Z}_{N}$. The property C2 is equivalent to that, for each $m \neq 0 \bmod N,\left(\pi_{i}(x), \pi_{i}(x+m)\right)=$ $\left(\pi_{j}(y), \pi_{j}(y+m)\right)$ if and only if $i=j$ and $x=y$, where $0 \leq i, j \leq r-1$ and $0 \leq x, y \leq N-1$. Circular Florentine rectangles mean that $x+m,y+m$ modulo $N$, and acyclic Florentine rectangles signify that $\pi_{i}(x+m)$ will be vanished if $x+m\geq N$.
The following matrix is an example of $4\times 15$ CFR.
	\begin{equation}\label{flo15}
		\left[\begin{array}{ccccccccccccccc}
			0 & 1 & 2 & 3 & 4 & 5 & 6 & 7 & 8 & 9 & 10 & 11 & 12 & 13 & 14 \\
			0 & 7 & 1 & 8 & 2 & 12 & 3 & 11 & 9 & 4 & 13 & 5 & 14 & 6 & 10 \\
			0 & 4 & 11 & 7 & 10 & 1 & 13 & 9 & 5 & 8 & 3 & 6 & 2 & 14 & 12 \\
			0 & 13 & 7 & 2 & 11 & 6 & 14 & 10 & 3 & 5 & 12 & 9 & 1 & 4 & 8
		\end{array}\right].
	\end{equation}
\end{definition}

For each positive integer $N\ge2$, we denote by
$\tilde{F}(N)$ the maximum number of rows such that an $\tilde{F}(N)\times N$ CFR exists.
Some known results of $\tilde{F}(N)$ are given in the following lemma.
\begin{lemma}[\cite{songF,thesisF}] For $N\geq 2$, we have the following bounds for $\tilde{F}(N)$:
	\begin{itemize}
		\item $ \tilde{F}(N)=1$ when $N$ is even,
		\item $p-1 \leq \tilde{F}(N) \leq N-1,$ where $p$ is the smallest prime
		factor of $N$,
		\item $\tilde{F}(N)=N-1$ when $N$ is a prime,
		\item $\tilde{F}(N) \leq N-3$ when $N \equiv 15 \bmod 18$.
	\end{itemize}
\end{lemma}

\begin{lemma}[\cite{thesisF}]\label{lemma2}
		Let $p$ be an odd prime integer. Then the multiplication table of $\mathbb{Z}_p$, without the upper all-zero row, is a $(p-1)\times p$ CFR. This also implies that for $N=p$, where $p$ is an odd prime, $\tilde{F}(N)=p-1$.
\end{lemma}

%
%
In this paper, we are interested in the bounds on $\tilde{F}(N)$, because they determine the set size of an asymptotically optimal set to be developed in a later part of in this paper. In Table \ref{posv}, some possible values of $\tilde{F}(N)$ are given based on the results given in \cite{songF}.

\begin{table}
	\small
	\centering
	\caption{Possible values of $\tilde{F}(N)$ for various odd-composite $N$\label{posv} \cite{songF}.}
	\begin{tabular}{|c|c|c|c|}
		\hline
		$N$ & \makecell{Possible value \\of $\tilde{F}(N)$} & $N$ & \makecell{Possible value \\of $\tilde{F}(N)$}  \\
		\hline
		9 &2 & 45 & $2, \dots,43$ \\
		\hline
		15 & 4 & 49 & $6, \dots,48$ \\
		\hline
		21 & $5, \dots, 19$ & 51 & $2, \dots,48$ \\
		\hline
		25 &  $4, \dots,24$ & 55&$4, \dots,54$ \\
		\hline
		27 & $4, \dots,26$ & 57 & $7, \dots,55$ \\
		\hline
		33 & $3, \dots,30$ & 63 & $6, \dots,62$ \\
		\hline
		35 & $4, \dots,33$ & 65 & $4, \dots,63$ \\
		\hline
		39 & $3, \dots,38$ & 69 & $2, \dots,66$\\
		\hline
	\end{tabular}
\end{table}

To proceed,
let us present the following lemmas which are useful for our
subsequent proof.

\begin{lemma}\label{lem4}
	Let $\mathcal{A}$ be an $M\times N$ CFR on $\mathbb{Z}_{N}$. Then each row, denoted by $\pi_{i}$ for $0 \le i \le M-1$, is a permutation of $\mathbb{Z}_{N}$. For $0 \leq i \neq r \leq M-1, \pi_{i}(t)=\pi_{r}\left(t+l^{\prime}\right)$ exactly has one solution for each $0 \leq l^{\prime} \leq N-1$.
\end{lemma}

\begin{IEEEproof}
	Based on the definition of CFR, for each $l \not\equiv 0 \bmod ~n$, we have $(\pi_{i}(j), \pi_{i}(j + l)) =(\pi_{r}(k), \pi_{r}(k + l))$ if and only if $i = r$ and $j = k$. Assume there exists $l^{\prime}$ such that $\pi_{i}(t)=\pi_{r}\left(t+l^{\prime}\right)$ for $i\neq r$ has two solutions $t_{1}$ and $t_{2}$. Then we obtain $\pi_{i}\left(t_{1}\right)=$ $\pi_{r}\left(t_{1}+l^{\prime}\right)$ and $\pi_{i}\left(t_{2}\right)=\pi_{r}\left(t_{2}+l^{\prime}\right) .$ Hence,
	$\left(\pi_{i}\left(t_{1}\right), \pi_{i}\left(t_{2}\right)\right)=\left(\pi_{r}\left(t_{1}+l^{\prime}\right), \pi_{r}\left(t_{2}+l^{\prime}\right)\right)$ for $i\neq r$, which leads to a contradiction. Therefore, we have $
\left|\left\{t: \pi_{i}(t)=\pi_{r}\left(t+l^{\prime}\right)\right\}\right|\leq1$ for each $0 \leq l^{\prime} \leq N-1$.
Since $\pi_{i}$ and $\pi_{r}$ are permutations of $\mathbb{Z}_{N}$, we have $
\sum_{l^{\prime}=0}^{N-1}\left|\left\{t: \pi_{i}(t)=\pi_{r}\left(t+l^{\prime}\right)\right\}\right|=N.
$ Hence, $\left|\left\{t: \pi_{i}(t)=\pi_{r}\left(t+l^{\prime}\right)\right\}\right|=1$ for each $0 \leq l^{\prime} \leq N-1$. Then the result follows.
\end{IEEEproof}

\begin{lemma}\label{lem5}
	In the context of the symbols in \textit{Lemma \ref{lem4}}, let $\pi^{-1}_{i}$ be the inverse of $\pi_{i}$ for $0 \le i \le M-1$, then $\pi^{-1}_{i}$ is a permutation of $\mathbb{Z}_{N}$, and $\pi^{-1}_{i}-\pi^{-1}_{r}$ is still a permutation of $\mathbb{Z}_{N}$ if $0 \leq i \neq r \leq M-1$.
\end{lemma}

\begin{IEEEproof}
	Obviously, $\pi^{-1}_{i}$ is a permutation of $\mathbb{Z}_{N}$ due to $\pi_{i}$ is a permutation of $\mathbb{Z}_{N}$. To prove $\pi^{-1}_{i}-\pi^{-1}_{r}$ is a permutation of $\mathbb{Z}_{N}$ is equivalent to prove
$$
\left(\pi^{-1}_{i}-\pi^{-1}_{r}\right)(x)\neq \left(\pi^{-1}_{i}-\pi^{-1}_{r}\right)(y), ~~\text{for~any}~~ 0\leq x\neq y\leq N-1.
$$
Suppose there exists a pair $(x,y)$ with $0\leq x\neq y\leq N-1$, such that $\left(\pi^{-1}_{i}-\pi^{-1}_{r}\right)(x)= \left(\pi^{-1}_{i}-\pi^{-1}_{r}\right)(y)$. Then we have $\pi^{-1}_{i}(x)-\pi^{-1}_{r}(x)\equiv\pi^{-1}_{i}(y)-\pi^{-1}_{r}(y)\equiv k\pmod N$ for an integer $k\in\mathbb{Z}_{n}$. Let $\pi^{-1}_{r}(x)=a,\pi^{-1}_{r}(y)=b$, we have
\begin{equation*}
\begin{split}
\left\{
\begin{array}{l}
\pi_{r}(a)=\pi_{i}(a+k)=x,\\
\pi_{r}(b)=\pi_{i}(b+k)=y.
\end{array}
\right.
\end{split}
\end{equation*}
Since $\pi^{-1}_{r}$ is a permutation and $x\neq y$, we have $a\neq b$. Hence, $\pi_{r}(t)=\pi_{i}\left(t+k\right)$ has two solutions. Which is contradict to \textit{Lemma \ref{lem4}}. The result then follows.

\end{IEEEproof}

\section{Strengthened Periodic Correlation Lower Bounds of SCSs}
Let $\mathcal{C}=\{C_0,C_1,\dots ,C_{M-1}\}$ be a set consisting of $M$ sequences of length $L$. Consider the integer time delay $\tau$ within the maximum
range of interest $L_{CZ}$. Define
\begin{equation}
	\begin{split}
		\theta_{a}(\mathcal{C})&=\max\{|\theta_{C_i}(\tau)|:0\leq i<M,0<\tau<L_{CZ}\},\\
		\theta_{c}(\mathcal{C})&=\max\{|\theta_{C_i,C_j}(\tau)|:0\leq i\neq j<M,0\leq \tau<L_{CZ}\}.
	\end{split}
\end{equation}
Then the maximum periodic correlation magnitude is defined as
\begin{equation}
	\theta_{\max}(\mathcal{C})=\max\{\theta_{a}(\mathcal{C}),\theta_{c}(\mathcal{C})\}.
\end{equation}
For multiple sets $\mathfrak{C}=\{\mathcal{C}^0,\mathcal{C}^1,\dots ,\mathcal{C}^{K-1}\}$, it can be similarly defined that $\theta_{a}(\mathfrak{C})=\max\{\theta_{\max}(\mathcal{C}^i):0\leq i<K\}$, $\theta_{c}(\mathfrak{C})=\max\{|\theta_{C^i_{l},C^j_{m}}(\tau)|:0\leq i\neq j<K,0\leq l, m<M,0\leq \tau<L_{CZ}\}$
and $\theta_{\max}(\mathfrak{C})=\max\{\theta_{a}(\mathfrak{C}),\theta_{c}(\mathfrak{C})\}$.
Numerous research on single SCS set, which can be seen as a special case of multiple SCS sets with each set containing one sequence, attempts over the past years.

In 2011, \cite{tsai} derived a relation between $\theta_{a}(\mathcal{C})$ and $\theta_{c}(\mathcal{C})$ as follows:
\begin{equation}\label{eq6}
	(L_{CZ}-1)\theta^2_{a}(\mathcal{C})+(M-1)L_{CZ}\theta^2_{c}(\mathcal{C})+L^{2}\geq \frac{ML^{2}L_{CZ}}{L-n}.
\end{equation}
where $|\Omega|=n$ (i.e., $\Omega$ contains $n$ elements).
In particular, let $\theta_{a}(\mathcal{C})=\theta_{c}(\mathcal{C})=0$, we obtain the tradeoff among
the parameters of any ZCZ SCS set below
\begin{equation}\label{eq10}
	(L-n)\geq MZ.
\end{equation}
By setting $L_{CZ}=L$, a generalized periodic correlation lower bound has been developed in \cite{liu2018}, which includes the zero power leakage as a special case as shown below:


\begin{lemma}\label{2018Liubound}
	Let $\mathcal{C}$ be an SCS family described as above, and $\theta_{\max}(\mathcal{C})$ be its maximum periodic correlation magnitude. Then
	\begin{equation}\label{Liubound}
		\theta_{\max}(\mathcal{C})\geq \theta_{opti}(\mathcal{C})= L\cdot \sqrt{\frac{(M-1)L+n}{(L-n)(ML-1)}}.
	\end{equation}
We call an SCS family $\mathcal{C}$ optimal if $\theta_{\max}(\mathcal{C})= \theta_{opti}(\mathcal{C})$.
An SCS family $\mathcal{C}$ is called asymptotically optimal if
$$
\lim_{L\rightarrow\infty}\frac{\theta_{\max}(\mathcal{C})}{\theta_{opti}(\mathcal{C})}= 1.
$$
\end{lemma}

Next, we propose an improved lower bound of $\theta_{a}$ and $\theta_{c}$ for SCSs. The key idea of the technique is to make full use of fact that uniform power allocation is adopted over all the admissible carriers.

\begin{theorem}\label{thbound}
For any $(M,L,\theta_{\max})$ SCS family $\mathcal{C}$, we have
\begin{equation*}\label{SC}
\begin{split}
\theta_{a}\geq L\sqrt{\frac{n}{(L-n)(L-1)}},~~\theta_{c}\geq \frac{L}{\sqrt{L-n}}.
\end{split}
\end{equation*}
\end{theorem}

\begin{IEEEproof}
For any $C_{i},C_{j}\in \mathcal{C}$, we have
\begin{equation}\label{sum}
\begin{split}
\sum_{\tau=0}^{L-1}\left|\theta_{C_{i},C_{j}}(\tau)\right|^{2}
&=\sum_{\tau=0}^{L-1}\left|\sum\limits_{f=0}^{L-1} \hat{c}_{i,f} \hat{c}_{j,f}^{*}  \omega^{-f\tau}_{L}\right|^{2}\\
&=\sum\limits_{f,f'=0}^{L-1} \hat{c}_{i,f} \hat{c}_{j,f}^{*}  \hat{c}_{i,f'}^{*}  \hat{c}_{j,f'} \sum_{\tau=0}^{L-1}\omega^{(f'-f)\tau}_{L}\\
&=L\sum\limits_{f=0}^{L-1} \left|\hat{c}_{i,f}\right|^{2}\left| \hat{c}_{j,f}\right|^{2} \\
&=\frac{L^{3}}{L-n}.
\end{split}
\end{equation}
Then the results follow from
\begin{equation*}\label{SC}
\begin{split}
\theta_{a}\geq \sqrt{\frac{\sum_{\tau=1}^{L-1}\left|\theta_{C_{i},C_{i}}(\tau)\right|^{2}}{L-1}},~~\theta_{c}\geq \sqrt{\frac{\sum_{\tau=0}^{L-1}\left|\theta_{C_{i},C_{j}}(\tau)\right|^{2}}{L}}.
\end{split}
\end{equation*}
\end{IEEEproof}
{
\begin{remark}
By \textit{Theorem \ref{thbound}}, adding the two inequalities, we have
\begin{equation}\label{newbound}
\theta_{a}^{2}(L-1)+\theta_{c}^{2}(M-1)L \geq \frac{L^{2}(ML-L+n)}{L-n}.
\end{equation}
The lower bound in (\ref{newbound}) is equivalent to $(12)$ in \cite{tsai}, i.e.,
$$
	(L_{CZ}-1)\theta^2_{a}+(M-1)L_{CZ}\theta^2_{c}+L^{2}\geq \frac{ML^{2}L_{CZ}}{L-n},
$$
when $Lcz=L$. (\ref{newbound}) is also equivalent to the classical Sarwate-bound if $n=0$. When we set $\theta_{a}=\theta_{c}=\theta_{\max}$, (\ref{newbound}) will be reduced to inequality $(44)$ in \cite{liu2018}, e.g.,
$$
\theta_{\max } \geq L \cdot \sqrt{\frac{(M-1)L+n}{(L-n)(ML-1)}}.
$$
\end{remark}}
{

Based on \textit{Theorem \ref{thbound}}, we obtain a lower bound of cross-correlation between different SCS families for multiple SCS sets as follow. To the best of our knowledge, \textit{Theorem \ref{thZCZ}} is the first lower bound of inter-set cross-correlation for multiple SCS sets.
}

\begin{theorem}\label{thZCZ}
Let $\mathfrak{C}=\{\mathcal{C}^0,\mathcal{C}^1,\dots ,\mathcal{C}^{K-1}\}$ be a set of $K$ SCS sets, each consisting of $M$ sequences of length $L$, then we have
\begin{equation}\label{cross-ZCZ}
\begin{split}
\theta_{c}(\mathfrak{C})\geq \frac{L}{\sqrt{L-n}}.
\end{split}
\end{equation}
\end{theorem}

%

\section{Proposed construction of Asymptotically Optimal SCS Using CFRs}
In this section, we propose a direct construction of a set of SCSs with asymptotically optimal correlation. These SCS sets have potential applications such as synchronization and channel estimation in cognitive networks.

\begin{construction}\label{cons1}
	Consider any odd positive integer $N\geq 2$, for which an $\tilde{F}(N)\times N$ Florentine rectangle $\mathcal{A}$ exists over $\mathbb{Z}_N$. Also let $\pi_k$ be a permutation over $\mathbb{Z}_N$ for $0\leq k <\tilde{F}(N)$, defined as above, which satisfies \textit{Lemma \ref{lem4}}. Let $\mathfrak{C}=\{\mathcal{C}^0,\mathcal{C}^1,\dots ,\mathcal{C}^{\tilde{F}(N)-1}\}$ be a set of $\tilde{F}(N)$ sequence sets, each containing a single sequence of length $L=N(N+1)$, i.e.,
	\begin{equation}
		\mathcal{C}^m=[c^m_0,c^m_1,\dots,c^m_{L-1}]_{1\times L}, ~0\leq m<\tilde{F}(N),
	\end{equation}
	where
	\begin{equation}\label{eq11}
		c^m_i=\omega^{\pi_m(\langle i\rangle_N) \cdot i}_{N+1},~ 0\leq i <L.
	\end{equation}
\end{construction}

For the sequence set generated in \textit{Construction \ref{cons1}} we have the following theorem.

\begin{theorem}\label{th1}
	$\mathfrak{C}$ described in \textit{Construction \ref{cons1}} is an SCS family over alphabet $\mathbb{Z}_{N+1}$ having the following properties:
	\begin{enumerate}
		\item $\theta_{a}(\mathfrak{C})=\theta_{c}(\mathfrak{C})=\theta_{\max}(\mathfrak{C})=N+1$.
		\item The spectral constraint set $\Omega$ for $\mathfrak{C}$ is $\Omega=\{1+a(N+1):a\in \mathbb{Z}_N\}$.
	\end{enumerate}
\end{theorem}

\begin{IEEEproof}
	By \textit{Construction \ref{cons1}}, $\mathfrak{C}$ contains $\tilde{F}(N)$ sequence sets each containing single sequence of length $L$. We prove the properties of $\mathfrak{C}$ as follows:
	\begin{enumerate}
		\item Let $\mathcal{C}^m$ and $\mathcal{C}^{m^\prime}$, for $0\leq m,m^{\prime}<\tilde{F}(N)$, be two sequences in $\mathfrak{C}$. We have
		\begin{equation}\label{eq12}
			\begin{split}
				\theta_{\mathcal{C}^m,\mathcal{C}^{m^\prime}}(\tau)&=\sum_{i=0}^{L-1}c^m_i \cdot (c^{m^\prime}_{i+\tau} )^*\\
				&=\sum_{i=0}^{N(N+1)-1}\omega^{\pi_m(\langle i\rangle_N) \cdot i-\pi_{m^\prime}(\langle i+\tau\rangle_N) \cdot (i+\tau)}_{N+1}\\
				&=\sum_{i_0=0}^{N-1}\omega_{N+1}^{-\pi_{m^\prime}(\langle i_0+\tau_0 \rangle_N)\cdot \tau_1}\cdot \sum_{i_1=0}^{N} \omega_{N+1}^{(\pi_m(\langle i_0 \rangle_N)-\pi_{m^\prime}(\langle i_0+\tau_0 \rangle_N))\cdot i_1},
			\end{split}
		\end{equation}
		where $i_0=\langle i \rangle_N$, $i_1=\langle i \rangle_{N+1}$, $\tau_0=\langle \tau \rangle_N$, and $\tau_1=\langle \tau \rangle_{N+1}$. We have the following cases.
		\begin{enumerate}
			\item[Case 1:] When $m=m^\prime$, $\tau_0=0$ and $\tau_1\neq 0$, (\ref{eq12}) becomes
			\begin{equation}
				\theta_{\mathcal{C}^m}(\tau)=(N+1)\cdot \sum_{t_0=0}^{N-1}\omega_{N+1}^{-\pi_m(\langle i_0 \rangle_N)\cdot \tau_1}.
			\end{equation}
			As per \textit{Lemma \ref{lem4}}, $\pi_m(\langle i_0 \rangle_N)$ is a permutation on $\mathbb{Z}_N$. Hence,
			\begin{equation}
				\sum_{t_0=0}^{N-1}\omega_{N+1}^{-\pi_m(\langle i_0 \rangle_N)\cdot \tau_1}=-\omega_{N+1}^{-N\cdot \tau_1}.
			\end{equation}
			Therefore, we have $|	\theta_{\mathcal{C}^m}(\tau)|=N+1$.
			\item[Case 2:] When $m=m^\prime$ and $\tau_0\neq 0$, (\ref{eq12}) becomes
			\begin{equation}
				\theta_{\mathcal{C}^m}(\tau)=\sum_{i_0=0}^{N-1}\omega_{N+1}^{-\pi_{m}(\langle i_0+\tau_0 \rangle_N)\cdot \tau_1}\cdot \sum_{i_1=0}^{N} \omega_{N+1}^{(\pi_m(\langle i_0 \rangle_N)-\pi_{m}(\langle i_0+\tau_0 \rangle_N))\cdot i_1}.
			\end{equation}
			Note that for $\tau_0\neq 0$, $(\pi_m(\langle i_0 \rangle_N)\neq \pi_{m}(\langle i_0+\tau_0 \rangle_N))$, since $\pi_m(i_0)$ is a permutation on $\mathbb{Z}_N$. Then
			\begin{equation}
				\sum_{i_1=0}^{N} \omega_{N+1}^{(\pi_m(\langle i_0 \rangle_N)-\pi_{m}(\langle i_0+\tau_0 \rangle_N))\cdot i_1}=0.
			\end{equation}
			Therefore, $|\theta_{\mathcal{C}^m}(\tau)|=0$.
			\item[Case 3:] When $m \neq m^\prime$, then from (\ref{eq12}) we have
			\begin{equation}
				\theta_{\mathcal{C}^m,\mathcal{C}^{m^\prime}}(\tau)=\sum_{i_0=0}^{N-1}\omega_{N+1}^{-\pi_{m^\prime}(\langle i_0+\tau_0 \rangle_N)\cdot \tau_1}\cdot \sum_{i_1=0}^{N} \omega_{N+1}^{(\pi_m(\langle i_0 \rangle_N)-\pi_{m^\prime}(\langle i_0+\tau_0 \rangle_N))\cdot i_1}.
			\end{equation}
			Recall that permutations $\pi_m$ and $\pi_{m^\prime}$ satisfy \textit{Lemma \ref{lem4}}. Hence, $\pi_m(\langle i_0 \rangle_N)=\pi_{m^\prime}(\langle i_0+\tau_0 \rangle_N)$ for any $0\leq \langle i_0 \rangle_N <N, 0\leq \langle i_0+\tau_0 \rangle_N<N$, $m\neq m^\prime$ has at most one solution. Therefore, if there is no solution, then $|\theta_{\mathcal{C}^m,\mathcal{C}^{m^\prime}}(\tau)|=0$. If there is one solution, say $i^\prime_0$, with $0\leq \langle i^\prime_0 \rangle_N <N,0\leq \langle i^\prime_0+\tau_0 \rangle_N < N$, then we have
			\begin{equation}
				\begin{split}
					\theta_{\mathcal{C}^m,\mathcal{C}^{m^\prime}}(\tau)=&(N+1)\cdot \omega_{N+1}^{-\pi_{m^\prime}(\langle i^\prime_0+\tau_0 \rangle_N)\cdot \tau_1}\\&+\sum_{\substack{i_0=0,\\i_0\neq i^\prime_0}}^{N-1}\omega_{N+1}^{-\pi_{m^\prime}(\langle i_0+\tau_0 \rangle_N)\cdot \tau_1}\cdot \sum_{i_1=0}^{N} \omega_{N+1}^{(\pi_m(\langle i_0 \rangle_N)-\pi_{m^\prime}(\langle i_0+\tau_0 \rangle_N))\cdot i_1}\\
					=&(N+1)\cdot \omega_{N+1}^{-\pi_{m^\prime}(\langle i^\prime_0+\tau_0 \rangle_N)\cdot \tau_1}.
				\end{split}
			\end{equation}
			Hence, $|\theta_{\mathcal{C}^m,\mathcal{C}^{m^\prime}}(\tau)|=N+1$. Observing the three cases above, we conclude that $\theta_{\max}(\mathfrak{C})=N+1$.
		\end{enumerate}
		\item We now show that the spectral constraint set $\Omega$ for $\mathfrak{C}$ is $\Omega=\{1+a(N+1):a\in \mathbb{Z}_N\}$.
		
		Let $\widehat{\mathcal{C}}^m$ for $0\leq m <\tilde{F}(N)$ be the frequency domain dual corresponding to the sequence sets ${\mathcal{C}}^m$ in $\mathfrak{C}$, where $\widehat{\mathcal{C}}^m=\widehat{C}^m=[\widehat{c}^m_{0},\widehat{c}^m_{1},\dots,\widehat{c}^m_{L-1}]=[c^m_0,c^m_1,\dots,c^m_{L-1}]\cdot \frac{1}{\sqrt{L}}\mathcal{F}_L$. Then, for $0\leq j <L$ we have
		\begin{equation}\label{eq19}
			\begin{split}
				\widehat{c}^m_{j}&=\frac{1}{\sqrt{L}}\sum_{i=0}^{L-1}c^m_i\cdot f_{i,j}\\
				&=\frac{1}{\sqrt{L}}\sum_{i=0}^{L-1}\omega^{\pi_m(\langle i\rangle_N) \cdot i}_{N+1}\cdot f_{i,j}\\
				&=\frac{1}{\sqrt{N(N+1)}}\sum_{i=0}^{N(N+1)-1}\omega^{\pi_m(\langle i\rangle_N) \cdot i}_{N+1}\cdot \omega_{N(N+1)}^{-ij}\\
				&=\frac{1}{\sqrt{N(N+1)}}\sum_{i_0=0}^{N-1}\sum_{i_1=0}^{N}\omega^{(N\cdot \pi_m(i_0) -j)\cdot(Ni_1+i_0)}_{N(N+1)}\\
				&=\frac{1}{\sqrt{N(N+1)}}\sum_{i_0=0}^{N-1}\omega^{(N\cdot \pi_m(i_0) -j)\cdot i_0}_{N(N+1)}\sum_{i_1=0}^{N}\omega^{(N\cdot \pi_m(i_0) -j)\cdot i_1}_{N+1},
			\end{split}
		\end{equation}
		where $i_0=\langle i \rangle_N$ and $i_1=\lfloor i/N \rfloor$. Since $\pi_m(i_0)$ for $0\leq i_0 <N$ is a permutation on $\mathbb{Z}_N$, we have $\{\langle N  \cdot \pi_m(i_0) \rangle_{(N+1)} : i_0 \in \mathbb{Z}_N \}=\mathbb{Z}_{N+1}\setminus \{1\}$. Therefore when $j\in \{1+a(N+1):a\in \mathbb{Z}_N\}$, one has $\langle (N\cdot \pi_m(i_0) -j) \rangle_{N+1}\neq 0$ for any $0\leq i_0 <N$. In this case $\widehat{c}^m_{j}=0$ holds for $\sum_{i_1=0}^{N}\omega^{(N\cdot \pi_m(i_0) -j)\cdot i_1}_{N+1}=0$ in (\ref{eq19}). Otherwise, there is only one solution $0\leq i^\prime_0<N$ such that $\langle (N\cdot \pi_m(i^\prime_0) -j) \rangle_{N+1}= 0$. Then, from (\ref{eq19}), we get
		\begin{equation}
			\begin{split}
				\widehat{c}^m_{j}&=\frac{1}{\sqrt{N(N+1)}}\omega^{(N\cdot \pi_m(i^\prime_0) -j)\cdot i^\prime_0}_{N(N+1)} \cdot (N+1)\\
				\implies |\widehat{c}^m_{j}|&=\sqrt{\frac{N+1}{N}}.
			\end{split}
		\end{equation}
		Therefore, $\sum_{m=0}^{K-1}|\widehat{c}^m_{j}|^2=0$ for all $j \in \Omega$, where $\Omega=\{1+a(N+1):a\in \mathbb{Z}_N\}$. Hence, $\mathfrak{C}$ is a SCS family with spectrall-null constraint $\Omega$.
	\end{enumerate}
\end{IEEEproof}
\begin{theorem}
	The proposed SCS family $\mathfrak{C}$ in \textit{Theorem \ref{th1}} is optimal with respect to the lower bound in \textit{Theorem \ref{thZCZ}}. Besides, the SCS family is asymptotically optimal with respect to (\ref{Liubound}) when $\lim\limits_{N\rightarrow\infty}\tilde{F}(N)=\infty.$
\end{theorem}
	
\begin{IEEEproof}
By \textit{Theorem \ref{thZCZ}}, we have
$$
\theta_{c}\geq \frac{N(N+1)}{\sqrt{N(N+1)-N}}=N+1.
$$
Hence, the SCS family $\mathfrak{C}$ has optimal cross-correlation. By (\ref{Liubound}), we have
		\begin{equation}
		\theta_{\max}(\mathfrak{C})\geq \theta_{opti}(\mathfrak{C})= L\cdot \sqrt{\frac{(K-1)L+n}{(L-n)(KL-1)}}.
	\end{equation}
	The resultant sequence family of our construction have $K=\tilde{F}(N)$, $n=N$, $L=N(N+1)$,

Denote the optimality factor by $\eta$, i.e.
\begin{equation}
	\eta=\frac{\theta_{\max}(\mathfrak{C})}{\theta_{opti}(\mathfrak{C})}.
\end{equation}
We have
\begin{equation}
	\begin{split}
		\eta&=\frac{N+1}{N(N+1)\sqrt{\frac{(\tilde{F}(N)-1)N(N+1)+N}{(N(N+1)-N)(\tilde{F}(N)N(N+1)-1)}}}\\
		&=\sqrt{\frac{\tilde{F}(N)N(N+1)-1}{(\tilde{F}(N)-1)N(N+1)+N}}\\
		&= \sqrt{1-\frac{(N+1)}{(\tilde{F}(N)-1)N(N+1)+N}+\frac{N(N+1)}{(\tilde{F}(N)-1)N(N+1)+N}}\\
		&= \sqrt{1-\frac{1}{(\tilde{F}(N)-1)N+\frac{1}{1+\frac{1}{N}}}+\frac{1}{(\tilde{F}(N)-1)+\frac{1}{(N+1)}}}.
	\end{split}
\end{equation}
So, if $\lim\limits_{N\rightarrow\infty}\tilde{F}(N)=\infty$, we have
\begin{equation}
	\lim\limits_{L\rightarrow \infty}\eta=1.
\end{equation}
Hence the resultant SCSs are asymptotically optimal. This completes the proof.
\end{IEEEproof}
\begin{corollary}
The proposed SCS family $\mathfrak{C}$ in \textit{Theorem \ref{th1}} is asymptotically optimal when
\begin{enumerate}
  \item $N=p$, where $p$ is prime.
  \item $N=p(p+k)$, where $p$ and $p+k$ are prime for some integers $k$.
  \item $N=\prod\limits_{i=1}^{h}\left(m+k_{i}\right)$, where $m+k_{1},\ldots,m+k_{h}$ are prime and $\mathcal{H}=\left\{k_{1}, \ldots, k_{h}\right\}$ of distinct nonnegative integers satisfies the `admissible' condition in \cite{maynard2015}.
\end{enumerate}
\end{corollary}
\begin{IEEEproof}
\begin{enumerate}
  \item When $N=p$, the result follows from \textit{Lemma \ref{lemma2}} and \textit{Theorem 2}.
  \item When $N=p(p+k)$, by Zhang's groundbreaking work \cite{zhang2014}, there exist infinite prime numbers $p$ such that $p,p+k$ are prime for some integers $k \leq7\times10^{7}$. Recently this bound has been tightened to $k\leq246$. Hence, there exists integer $k\leq246$ meeting the condition. Specially, if the twin-prime conjecture holds, we can choose $k=2$. Then we have
      $
      \lim\limits_{N\rightarrow\infty}\tilde{F}(N)=\sqrt{N}=\infty,
      $
      the result follows from \textit{Theorem 2}.
  \item When $N=\prod_{i=1}^{h}\left(m+k_{i}\right)$, by Maynard's work \cite{maynard2015}, the prime $k$-tuples conjecture holds for a positive proportion of admissible $k$-tuples. Hence, there exist infinite positive numbers $m$ such that $m+k_{i}$ are prime. Then we have
      $
      \lim\limits_{N\rightarrow\infty}\tilde{F}(N)=\sqrt[h]{N}=\infty.
      $
      The result then follows from \textit{Theorem 2}.
\end{enumerate}

\end{IEEEproof}
\begin{remark}
	Since systematic constructions of CFRs for all matrix orders are not available and the availability of CFRs are highly based on computer search results, we calculate the maximum value of the optimality factor $\eta$ in Table \ref{neweta}, for the available values of $\tilde{F}(N)$ under different values of $N$, given in Table \ref{posv}.
	\begin{table}
		\centering
		\caption{Possible values of $\tilde{F}(N)$ for various odd-composite $N$\label{neweta}.}
		\begin{tabular}{|c|c|c|c|c|c|c|}
			\hline
			$N$ & Length of seq. & $\tilde{F}(N)$ & $\theta_{\max }$ & $\theta_{\text {opti }}$ & $\eta$ & $\eta$ in \cite{Tian2020}\\
			\hline
			15 & 240 & $\geq4$ & 16 & $\geq14.0073$ & $\leq1.1423$& $1.3706$\\
			21 & 462 & $\geq5$ & 22 & $\geq19.7932$ & $\leq1.1115$& $1.3824$\\
			25&	650	&$\geq4$	&26	&$\geq22.6649$ 	&$\leq1.1471$&1.1471\\
			27 & 756 & $\geq4$ & 28 & $\geq24.3967$ & $\leq1.1477$& $1.3892$\\
			33 & 1122 & $\geq3$ & 34 & $\geq27.9684$ & $\leq1.2157$& $1.3936$\\
			35&	1260	&$\geq4$	&36	&$\geq31.3240$ 	&$\leq1.1493$ &1.1493\\
			39 & 1560 & $\geq3$ & 40 & $\geq32.8669$ & $\leq1.2170$& $1.3966$\\
			45&	2070&	$\geq2$&	46&	$\geq32.8825$& 	$\leq1.3989$& 	1.3989\\
			49&	2450&	$\geq6$&	50&	$\geq45.7363$& 	$\leq1.0932$& 	1.0932\\
			51&	2652&	$\geq2$&	52&	$\geq37.1249$& 	$\leq1.4007$& 	1.4007\\
			55&	3080&	$\geq4$&	56&	$\geq48.6435$& 	$\leq1.1512$& 	1.1512\\
			57 & 3306 & $\geq7$ & 58 & $\geq53.7758$ & $\leq1.0786$& $1.4021$\\
			63 & 4032 & $\geq6$ & 64 & $\geq58.5162$ & $\leq1.0937$& $1.4032$\\
			69&	4830&	$\geq2$&	70&	$\geq49.8524$& 	$\leq1.4041$& 	1.4041\\
			\hline
		\end{tabular}
	\end{table}
	As we can see in Table \ref{neweta}, the minimum possible values of $\tilde{F}(N)$ hugely improves the value of the optimality factor $\eta$, compared to the previous results available in \cite{Tian2020}.
In particular, when $N$ is prime, we have $\tilde{F}(N)=p-1$ and the $\theta_{\max}(\mathfrak{C})$ is extremely close to the lower bound in \textit{Lemma \ref{2018Liubound}}.
\end{remark}

\begin{example}\label{exm1}
	Consider $N=15$. By using the $4\times 15$ CFR given in (\ref{flo15}) in \textit{Construction \ref{cons1}}, we get $\mathfrak{C}=\{\mathcal{C}^0,\mathcal{C}^1,\mathcal{C}^2,\mathcal{C}^3\}$, where each $\mathcal{C}^m$ for $0\leq m \leq 3$, consists of a single sequence of length $240$ and the elements are defined as per (\ref{eq11}). A glimpse of the autocorrelation and cross-correlation of the generated sequences in time domain in shown is the first two sub-plots of Fig. \ref{fig1}. As shown in Fig. \ref{fig1}, $\theta_{\max}(\mathfrak{C})=16$.
	
	Let $\widehat{\mathcal{C}}^m$ be the frequency domain dual of the sequence set $\mathcal{C}^m$ for $0\leq m \leq 3$. Here the magnitudes of $\widehat{c}^m_j$ for $0\leq m \leq 3$ are
	\begin{equation}
		|\widehat{c}^m_j|=\begin{cases}
			0 &\text{ for } j\in \Omega,\\
			\sqrt{\frac{16}{15}} & \text{ for } j\notin \Omega,
		\end{cases}
	\end{equation}
where $\Omega=\{ 1,~    17,~    33,~    49,~    65,~    81,~    97,~   113,~   129,~   145,~   161,~   177,~   193 ,~  209,~   225\}$ is the spectral-null constraint. The third sub-plot of Fig. \ref{fig1} shows a glimpse of the magnitudes of $\widehat{c}^m_j$. Hence, $\sum_{m=0}^{3}|\widehat{c}^m_{j}|^2=0$ for all $j \in \Omega$. Therefore, $\mathfrak{C}$ is an SCS family with spectrall-null constraint $\Omega$.

	\begin{figure}
		\includegraphics[draft=false,width=\textwidth]{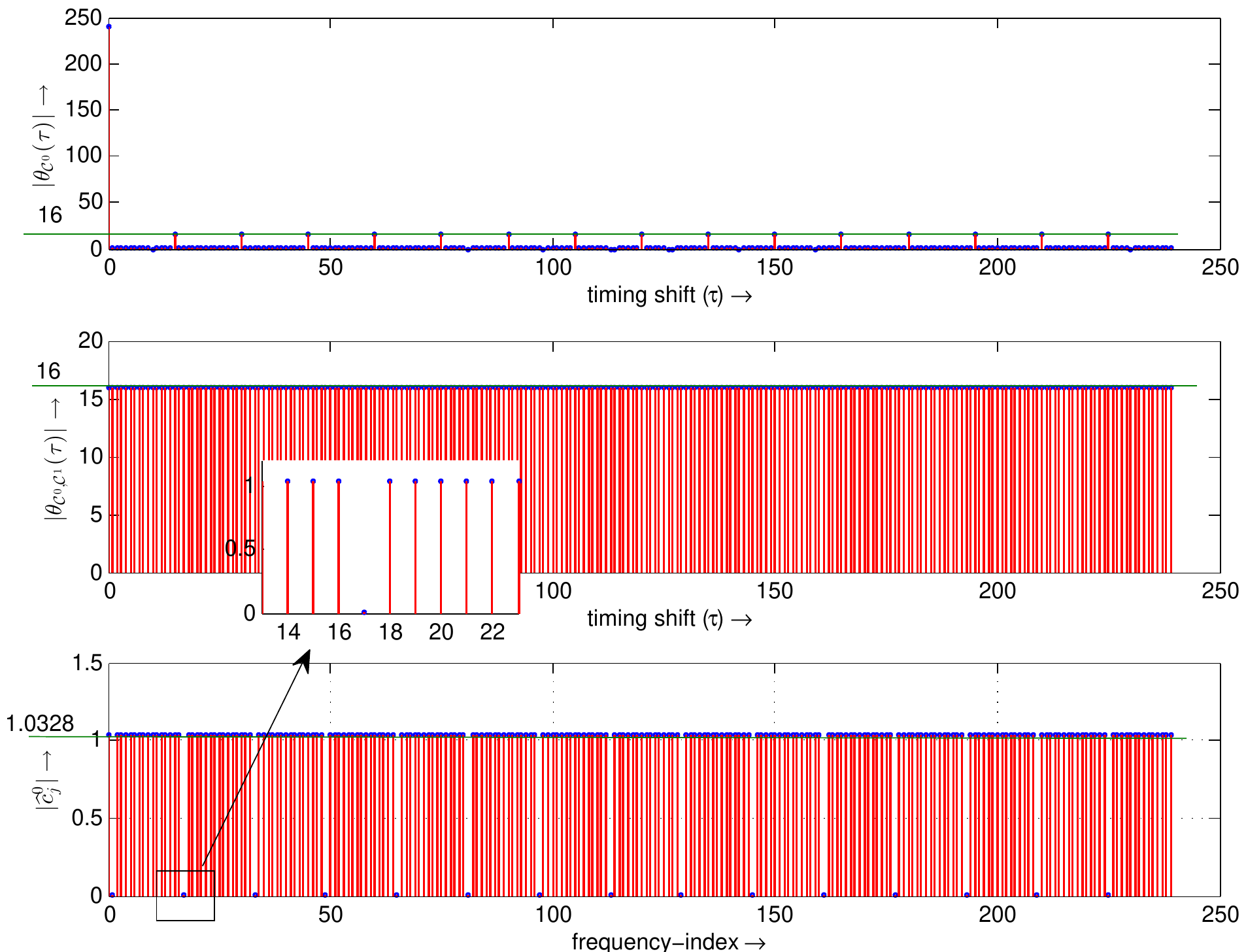}
		\caption{Illustrative plots of periodic auto- and cross- correlation magnitudes and a frequency-domain dual of the SCS family developed in \textit{Example \ref{exm1}}.}\label{fig1}
	\end{figure}
\end{example}
	
\section{Proposed unifying framework of SCS Sets Based on Interleaving Techniques}	
	In the constructions reported in \cite{Tian2020} and the sequences reported in \textit{Construction \ref{cons1}}, for a length $N(N+1)$ sequence, forbidden carrier positions are drawn from the set $\Omega=\{1+a(N+1):a\in \mathbb{Z}_N\}$. In this section, we will introduce an important construction framework, with which we can have more flexible forbidden  carrier positions. Besides, multiple SCS sets with ZCZ can also be obtained by the framework.

	
\subsection*{Proposed Unifying Framework}

\begin{itemize}
  \item Construct $M$ matrices $\mathcal{A}^i,~ 0\leq i <M$, of order $N \times N$, as follows
	\begin{equation}\label{eq30}
		\mathcal{A}^i=\begin{bmatrix}
			a^i_{0,0} & a^i_{0,1} & \dots & a^i_{0,N-1}\\
			a^i_{1,0} & a^i_{1,1} & \dots & a^i_{1,N-1}\\
			\vdots & \vdots & \ddots & \vdots \\
			a^i_{N-1,0} & a^i_{N-1,1} & \dots & a^i_{N-1,N-1}
		\end{bmatrix}_{N\times N},
	\end{equation}
{where the elements of this matrix $\mathcal{A}^i$ will be carefully designed in the sequel.}
	  \item Let $P=N+T$ where $T$ is a positive integer. Assuming $\mathcal{Z}=\mathbf{0}_{N\times 1}$ is inserted at the $s_i$-th column, $i=0,1,\cdots,T-1$, and all $s_i$ form a subset $\mathcal{I}$ of $\mathbb{Z}_{P}$, {obviously $T=\left|\mathcal{I}\right|$}. Define $\mathbb{Z}_{P}\setminus\mathcal{I}=\{l_{0},\cdots,l_{N-1}\}$ with $l_{0}<\cdots<l_{N-1}$, the $i$-th base matrix $\mathcal{B}^i$ of order $N \times P$ is given by $\mathcal{B}^i =\left(b^i_{j,k}\right)_{N\times P}$, where
	\begin{equation} \label{insertzero}
		b^i_{j,k}=\begin{cases}
			a^i_{j,t}, & \text{ if }k=l_{t};\\
			0, & \text{ if }k\in \mathcal{I}.
		\end{cases}
	\end{equation}
  \item Let $\widehat{U}^i=[\widehat{u}^{i}_{0},\widehat{u}^{i}_{1},\cdots,\widehat{u}^{i}_{L-1}]$ be the interleaved sequences of $\mathcal{B}^i$ with length $NP$ in the frequency domain, {i.e.,
       	\begin{equation}\label{eq32}
		\hat{u}^i_n=b^i_{r,s} \text{ for }r=\left\lfloor\frac{n}{P}\right\rfloor,s=n-Pr \text{ and }0\leq n <NP.
	\end{equation}}
       Then the sequence set $\mathcal{U}=\{U^0,U^1,\dots,U^{M-1}\}$, the collection of corresponding time-domain sequence of $\widehat{U}^i$, is an SCS set, having the following properties:
	\begin{enumerate}
		\item The spectral constraint set $\Omega=\{s+aP:s\in \mathcal{I},a\in \mathbb{Z}_N\}$.
		\item The auto-correlation of each SCS $U^i$ is given as
\begin{equation}\label{eqac}
  \theta_{U^{i}} (\tau)=\left\{
                         \begin{array}{ll}
                           NP, & \tau=0; \\
                           P\sum\limits_{l\in \mathbb{Z}_{P}\setminus\mathcal{I}}\omega_{P}^{lb}, &\tau=Nb ~{\rm{with}}~P\nmid b; \\
                           0, & \hbox{otherwise.}
                         \end{array}
                       \right.
  \end{equation}
		\item The cross-correlation between $U^{i_{1}}$ and $U^{i_{2}}$ is given as follow
\begin{equation}\label{eq35}
\begin{split}
	\theta_{U^{i_1}, U^{i_2}}(\tau)&=\sum\limits_{n=0}^{NP-1} \hat{u}^{i_1}_{n} \left(\hat{u}^{i_2}_{n}\right)^* \omega^{n\tau}_{NP}\\
	&= \sum\limits_{r=0}^{N-1}\sum\limits_{s=0}^{P-1} \hat{u}^{i_1}_{Pr+s} \left(\hat{u}^{i_2}_{Pr+s}\right)^* \omega^{(Pr+s)\tau}_{NP}\\
	&=\sum\limits_{s=0}^{P-1} \omega_{NP}^{ s\tau} \sum\limits_{r=0}^{N-1}b^{i_1}_{r,s} {b^{i_2}_{r,s}}^*\omega_{N}^{ r\tau}.
\end{split}
\end{equation}
        \item The element of time domain sequence is given as
        \begin{equation}\label{phase}
  u^{i}_{t}=\frac{1}{\sqrt{L}}\sum_{n=0}^{L-1}\hat{u}^{i}_{n}\omega_{L}^{nt}= \frac{1}{\sqrt{L}}\sum_{s=0}^{P-1}  \omega_{NP}^{st} \sum_{r=0}^{N-1} b^{i}_{r,s}\omega_{N}^{rt}.
  \end{equation}
\end{enumerate}

\end{itemize}
Note that, SCS sets generated by the above framework are capable of supporting more flexible spectral holes, whereas the correlations and element magnitudes are dependent on the base matrices. By choosing appropriate base matrices, some unimodular SCS sets with optimal correlation properties will be presented subsequently. In addition, we design optimal multiple ZCZ SCS sets by choosing special base matrices.
\section{Some optimal SCS sets derived from the unifying framework}	

\subsection{SCS Sets with Flexible Spectrum Constraints}
\begin{construction}\label{const2}
	Given an $M\times N$ CFR $\mathcal{F}$, let us consider an $M\times N$ matrix $\mathcal{G}=(g_{i,j})_{M\times N}$, where $g_{i,j}=\pi_{i}^{-1}(j)$ and $\pi_{i}$ is the $i$-th row of $\mathcal{F}$. {Let $P=N+1$ and $L=PN$, which is a special case for the proposed unifying framework with $T=1$.}
Define
	\begin{equation}\label{eq29}
		a^i_{j,k}=\sqrt{\frac{P}{N}}\omega_{N}^{jg_{i,k}}.
	\end{equation}
By using our framework, we have
	\begin{equation}\label{eq31}
		b^i_{j,k}=\begin{cases}
			a^i_{j,k}\omega_{NP}^{kg_{i,k}} & \text{ if }k<s_0;\\
			0 & \text{ if }k=s_0;\\
			a^i_{j,k-1}\omega_{NP}^{kg_{i,k-1}} & \text{ if }k>s_0.
		\end{cases}
	\end{equation}
	Using $M$ base matrices $\mathcal{B}^i$, let us construct $M$ interleaved sequences $\widehat{U}^i=[\hat{u}^i_0,\hat{u}^i_1,\dots,\hat{u}^i_{NP-1}]$ which is defined as equation (\ref{eq32}). For the corresponding time domain sequences of $\widehat{U}^i$, we have the following theorem.
\end{construction}	

\begin{theorem}\label{th2}
	The sequence set $\mathfrak{U}=\{\mathcal{U}^0,\mathcal{U}^1,\dots,\mathcal{U}^{M-1}\}$ defined in \textit{Construction 2}, where $\mathcal{U}^i=U^i$, the corresponding time-domain sequence of $\widehat{U}^i$, is an unimodular SCS with alphabet size less than $P$, having the following properties:
	\begin{enumerate}
	\item$\theta_{a}(\mathfrak{U})=\theta_{c}(\mathfrak{U})=\theta_{\max}(\mathfrak{U})=N+1$.
		\item The spectral constraint set $\Omega=\{s_0+a(N+1):a\in \mathbb{Z}_N\}$.
	\end{enumerate}
\end{theorem}
\begin{IEEEproof}
	First, let us prove the second part of the theorem. Since we have assumed that for $0\leq i <M$ the base {matrix} $\mathcal{B}^i$ have zero column $\mathcal{Z}$ at the $s_0$-th column for each $i$. Therefore, as per the proposed construction, the constructed interleaved sequences $\widehat{U}^i$ corresponding to the base {matrix} $\mathcal{B}^i$ will have zero at the positions $Pr+s_0,~0\leq r <N,~0\leq s_0<P$. In other words, for each $i$ with $0\leq i < M$,
	\begin{equation}
		\hat{u}^i_{(N+1)r+s_0}=0.
	\end{equation}
Therefore, $\sum_{i=0}^{M-1}|\hat{u}^i_j|^2=0$ for all $j\in \Omega$, where $\Omega=\{s_0+a(N+1):a\in \mathbb{Z}_N\}$. Hence, $\mathcal{U}$ is an SCS with spectral-null constraint $\Omega$. For all $0\leq i <M$, when $j\not\in \Omega$, we also have from (\ref{eq29}) and (\ref{eq31}),
\begin{equation}
	|\hat{u}^i_j|=\sqrt{\frac{P}{N}}=\sqrt{\frac{N+1}{N}}.
\end{equation}

Next, we prove the first part of the theorem. Let $0\leq i_1,i_2 <M$ and $U^{i_1}$ and $U^{i_2}$ be two time domain sequences corresponding to the frequency domain sequences $\widehat{U}^{i_1}$ and $\widehat{U}^{i_2}$, respectively. Then we have for $0\leq \tau <NP$
\begin{equation}\label{eq35}
\begin{split}
	\theta_{U^{i_1}, U^{i_2}}(\tau)&=\sum\limits_{n=0}^{NP-1} \hat{u}^{i_1}_{n} \left(\hat{u}^{i_2}_{n}\right)^* \omega^{n\tau}_{NP}\\
	&= \sum\limits_{r=0}^{N-1}\sum\limits_{s=0}^{P-1} \hat{u}^{i_1}_{Pr+s} \left(\hat{u}^{i_2}_{Pr+s}\right)^* \omega^{(Pr+s)\tau}_{NP}\\
	&=\sum\limits_{s=0}^{P-1} \omega_{NP}^{ s\tau} \sum\limits_{r=0}^{N-1}b^{i_1}_{r,s} {b^{i_2}_{r,s}}^*\omega_{N}^{ r\tau}\\
	&=\sum\limits_{s=0}^{s_0-1} \omega_{NP}^{ s\tau} \sum\limits_{r=0}^{N-1}b^{i_1}_{r,s} {b^{i_2}_{r,s}}^*\omega_{N}^{ r\tau}+\sum\limits_{s=s_0+1}^{P-1} \omega_{NP}^{ s\tau} \sum\limits_{r=0}^{N-1}b^{i_1}_{r,s-1} {b^{i_2}_{r,s-1}}^*\omega_{N}^{ r\tau}.
\end{split}
\end{equation}
We have the following cases. For $0\leq i_1=i_2<M$ and $\tau=0$, we have from (\ref{eq35}) and (\ref{eq29})
\begin{equation}\label{G1}
	\begin{split}
		&\theta_{U^{i_1}}(0)\\&=\sum\limits_{s=0}^{s_0-1}  \sum\limits_{r=0}^{N-1}a^{i_1}_{r,s} {a^{i_1}_{r,s}}^*+\sum\limits_{s=s_0+1}^{P-1}  \sum\limits_{r=0}^{N-1}a^{i_1}_{r,s-1} {a^{i_1}_{r,s-1}}^*\\
		&=\frac{P}{N}\left[\sum\limits_{s=0}^{s_0-1}  \sum\limits_{r=0}^{N-1}\omega_{NP}^{Pr(g_{i_1,s}-g_{i_1,s})+s(g_{i_1,s}-g_{i_1,s})}+\sum\limits_{s=s_0+1}^{P-1}  \sum\limits_{r=0}^{N-1}\omega_{NP}^{Pr(g_{i_1,s-1}-g_{i_1,s-1})+s(g_{i_1,s-1}-g_{i_1,s-1})}\right]\\
		&=\frac{P}{N}\left[s_0N+(N-s_0)N\right]=PN.
	\end{split}
\end{equation}
For $0\leq i_1=i_2<M$ and $0<\tau<NP$, we have from (\ref{eq35}) and (\ref{eq29})
\begin{equation}\label{eq37}
	\begin{split}
		&\theta_{U^{i_1}}(\tau)\\&=\sum\limits_{s=0}^{s_0-1} \omega_{NP}^{ s\tau} \sum\limits_{r=0}^{N-1}a^{i_1}_{r,s} {a^{i_1}_{r,s}}^*\omega_{N}^{ r\tau}+\sum\limits_{s=s_0+1}^{P-1} \omega_{NP}^{ s\tau} \sum\limits_{r=0}^{N-1}a^{i_1}_{r,s-1} {a^{i_1}_{r,s-1}}^*\omega_{N}^{ r\tau}\\
		&=\frac{P}{N}\left[\sum\limits_{s=0}^{s_0-1}\omega_{NP}^{ s\tau}  \sum\limits_{r=0}^{N-1}\omega_{NP}^{Pr(g_{i_1,s}-g_{i_1,s})+s(g_{i_1,s}-g_{i_1,s})}\omega_{N}^{ r\tau}\right. \\& \left. \hspace{2cm} +\sum\limits_{s=s_0+1}^{P-1}\omega_{NP}^{ s\tau}  \sum\limits_{r=0}^{N-1}\omega_{NP}^{Pr(g_{i_1,s-1}-g_{i_1,s-1})+s(g_{i_1,s-1}-g_{i_1,s-1})}\omega_{N}^{ r\tau}\right].
	\end{split}
\end{equation}
When $\tau=bN$ for some integer $b$, i.e., when $\tau <NP$ is a multiple of $N$, then we have from (\ref{eq37})
\begin{equation}\label{G2}
	\begin{split}
		|\theta_{U^{i_1}}(\tau)|&=\frac{P}{N}\left[N\left(\sum\limits_{s=0}^{s_0-1} \omega_{NP}^{ s\tau} +\sum\limits_{s=s_0+1}^{P-1} \omega_{NP}^{ s\tau}\right) \right]\\
		&=P|\omega_{P}^{ s_0b}|=P.
	\end{split}
\end{equation}
For other non-zero $\tau$, since $\sum\limits_{r=0}^{N-1}\omega_{N}^{ r\tau}=0$, we have from (\ref{eq37}) $|\theta_{U^{i_1}}(\tau)|=0$.

Next, for $0\leq i_1\neq i_2<M$, we have from (\ref{eq35}) and (\ref{eq29})
\begin{equation}\label{eq39}
	\begin{split}
		&\theta_{U^{i_1}, U^{i_2}}(\tau)\\&=\frac{P}{N}\left[\sum\limits_{s=0}^{s_0-1}\omega_{NP}^{ s\tau}  \sum\limits_{r=0}^{N-1}\omega_{NP}^{Pr(g_{i_1,s}-g_{i_2,s})+s(g_{i_1,s}-g_{i_2,s})}\omega_{N}^{ r\tau}\right. \\& \left. \hspace{2cm} +\sum\limits_{s=s_0+1}^{P-1}\omega_{NP}^{ s\tau}  \sum\limits_{r=0}^{N-1}\omega_{NP}^{Pr(g_{i_1,s-1}-g_{i_2,s-1})+s(g_{i_1,s-1}-g_{i_2,s-1})}\omega_{N}^{ r\tau}\right]\\
		&=\frac{P}{N}\left[\sum_{s=0}^{s_{0}-1}\right. \omega_{N P}^{s\left(g_{i_{1}, s}-g_{i_{2}, s}+\tau\right)} \sum_{r=0}^{N-1}  \omega_{N }^{ r\left(g_{i_{1}, s}-g_{i_{2}, s}+\tau\right)}  \\
		&\left.\hspace{2cm} +\sum_{s=s_{0}}^{N-1} \omega_{N P}^{(s+1)\left(g_{i_{1}, s}-g_{i_{2}, s}+\tau\right)} \sum_{r=0}^{N-1} \omega_{N }^{ r\left(g_{i_{1}, s}-g_{i_{2}, s}+\tau\right)} \right].
	\end{split}
\end{equation}
By \textit{Lemma \ref{lem5}}, we assert that $\pi^{-1}_{i_1}-\pi^{-1}_{i_2}$ is a permutation of $\mathbb{Z}_{N}$. Hence, $g_{i_{1}, s}-g_{i_{2}, s}+\tau\equiv0 \pmod N$ exactly has one solution for $s\neq s_{0}$, say $s^{\prime}$. If $g_{i_{1}, s}-g_{i_{2}, s}+\tau \not\equiv0 \pmod N$, then $\sum\limits_{r=0}^{N-1}\omega_{N}^{r(g_{i_1,s}-g_{i_2,s})}=0$. Hence, if $s^\prime <s_0$, then we have from (\ref{eq39})
\begin{equation}
		|\theta_{U^{i_1}, U^{i_2}}(\tau)|=\frac{P}{N}\left[N +0\right]=P.
\end{equation}
Similarly, when $s^\prime > s_0$, we have
\begin{equation}
	|\theta_{U^{i_1}, U^{i_2}}(\tau)|=\frac{P}{N}\left[0+N  \right]=P.
\end{equation}
Hence, observing the cases above, we conclude that $\theta_{\max}(\mathfrak{U})=P=N+1$.

In the end, we prove $U^{i}$ is a unimodular sequence. By the inverse Fourier transform, we have
\begin{equation}\label{phase}
\begin{split}
  &u^{i}_{t}=\frac{1}{\sqrt{L}}\sum_{n=0}^{L-1}\hat{u}^{i}_{n}\omega_{L}^{nt}= \frac{1}{\sqrt{L}}\sum_{s=0}^{P-1}  \omega_{NP}^{st} \sum_{r=0}^{N-1} b^{i}_{r,s}\omega_{N}^{rt}\\
  &=\frac{1}{\sqrt{L}}\left[\sum\limits_{s=0}^{s_0-1} \omega_{NP}^{ st} \sum\limits_{r=0}^{N-1}a^{i}_{r,s}\omega_{N}^{ rt}+\sum\limits_{s=s_0+1}^{P-1} \omega_{NP}^{ st} \sum\limits_{r=0}^{N-1}a^{i}_{r,s-1}\omega_{N}^{ rt}\right] \\
  	&=\frac{1}{N}\left[\sum_{s=0}^{s_{0}-1}\right. \omega_{N P}^{s\left(g_{i, s}+t\right)} \sum_{r=0}^{N-1}  \omega_{N }^{ r\left(g_{i, s}+t\right)} \left. +\sum_{s=s_{0}}^{N-1} \omega_{N P}^{(s+1)\left(g_{i, s}+t\right)} \sum_{r=0}^{N-1} \omega_{N }^{ r\left(g_{i, s}+t\right)} \right].
 \end{split}
\end{equation}
Based on \textit{Lemma \ref{lem5}}, $g_{i, s}+t\equiv 0 \pmod N$ exactly has one solution for $s\in\mathbb{Z}_{N}$. Hence, $|u^{i}_{t}|=1$ for any $t\in\mathbb{Z}_{L}$. In addition, the phase of $U^{i}$ is at most $P$ since $N\mid \left(g_{i, s}+t\right)$.
This completes the proof.

\end{IEEEproof}


\begin{example}\label{exm2}
	Let $\mathcal{F}$ be the $4\times 15$ CFR given in (\ref{flo15}). Here $M=4$, $N=15$, $P=N+1=16$, and $L=NP=240$. Let us choose $0\leq s_0<P=16$, arbitrarily, say, $s_0=7$. Then, according to \textit{Construction \ref{const2}}, we get $\mathfrak{U}=\{\mathcal{U}^0,\mathcal{U}^1,\dots ,\mathcal{U}^{3}\}$, where each of the $\mathcal{U}^i$ for $0\leq i <4$, consists of a single sequence of length $240$, constructed by interleaving the base {matrix} $\mathcal{B}^i$ of order $15\times 16$, as described in our framework. A glimpse of the autocorrelation and cross-correlation of the generated sequences in time domain is shown in the first two sub-plots of Fig. \ref{fig2}. As shown in Fig. \ref{fig2}, $\theta_{\max}(\mathfrak{C})=16$.
	
	As described in (\ref{eq32}), the spectrum of the elements $\hat{u}^i_n$ of the frequency domain sequences $\widehat{U}^i$ for $0\leq i<4$ are
	\begin{equation}
		|\hat{u}^i_n|=\begin{cases}
			0 &\text{ for } n\in \Omega,\\
			\sqrt{\frac{16}{15}} & \text{ for } n\notin \Omega,
		\end{cases}
	\end{equation}
where $\Omega=\{  7,~    23,~    39,~    55,~    71,~    87,~   103,~   119,~   135,~   151,~   167,~   183,~   199,~   215,~   231\}$ is the spectral-null constraint. The third sub-plot of Fig. \ref{fig2} shows a glimpse of the magnitudes of $u^i_n$. Hence, $\sum_{i=0}^{3}|u^i_n|^2=0$ for all $n \in \Omega$. Therefore, $\mathfrak{U}$ is an SCS family with spectral-null constraint $\Omega$.

	\begin{figure}
	\includegraphics[draft=false,width=\textwidth]{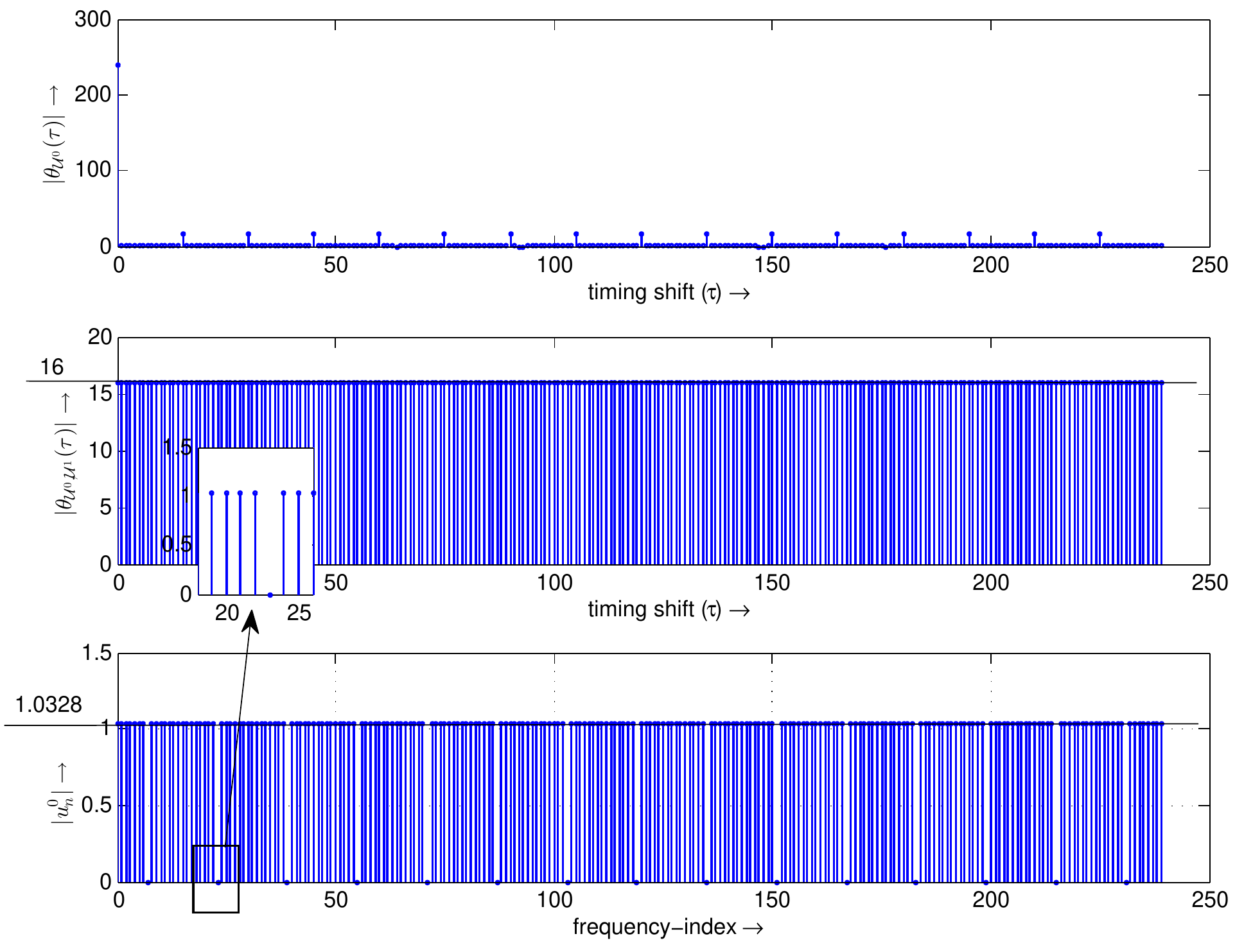}
	\caption{Illustrative plots of periodic auto- and cross- correlation magnitudes and a frequency-domain dual of the SCS family developed in
\textit{Example \ref{exm2}}.}\label{fig2}
\end{figure}
\end{example}

Based on the above discussion and the lower bound in \textit{Lemma \ref{2018Liubound}}, the correlation properties of the designed SCS family are closely related to spectral-null constraint $\Omega$. In fact, one can insert more spectral hole at a slight sacrifice of sequence autocorrelation. Next we present a generalization of \textit{Construction 2}, which provide more flexible spectral-null constraint $\Omega$.

\begin{construction}\label{gcon}(\textbf{Generalization of \textit{Construction 2}})	{Let $P=N+T$ where $T$ is a positive integer and the elements $a^{i}_{jk}$ of $\mathcal{A}^{i}$ are defined by (\ref{eq29}).} Define
	\begin{equation}
		b^i_{j,k}=\begin{cases}
			a^i_{j,t}\omega_{NP}^{kg_{i,t}} & \text{ if }k=l_{t};\\
			0 & \text{ if }k\in \mathcal{I},
		\end{cases}
	\end{equation}
{where $l_{t},\mathcal{I}$ are defined in equation (\ref{insertzero}) and $T$ is the cardinality of $\mathcal{I}$.}
Similarly, by our framework, we have the following theorem.
\end{construction}
\begin{theorem}
The sequence set $\mathfrak{U}=\{\mathcal{U}^0,\mathcal{U}^1,\dots,\mathcal{U}^{M-1}\}$ defined in \textit{Construction \ref{gcon}}, where $\mathcal{U}^i=U^i$, the corresponding time-domain sequence of $\widehat{U}^i$, is an SCS, having the following properties:
	\begin{enumerate}
		\item $\theta_{a}(\mathfrak{U})\geq P\sqrt{\frac{N(P-N)}{P-1}}$. If $\mathbb{Z}_{P}\setminus\mathcal{I}$ is a cyclic difference set over $\mathbb{Z}_{P}$, the equality holds.
        \item $\theta_{c}(\mathfrak{U})=P$.
		\item The spectral constraint set $\Omega=\{s+aP:s\in \mathcal{I},a\in \mathbb{Z}_N\}$.
	\end{enumerate}
\end{theorem}
 \begin{IEEEproof}
  The {proofs} of properties $2)$ and $3)$ are similar to that for \textit{Theorem \ref{th2}}, hence we only give a proof of property $1)$. Based on equation $(\ref{eqac})$, we have
  \begin{equation}
  \theta_{U^{i}} (\tau)=\left\{
                         \begin{array}{ll}
                           NP, & \tau=0; \\
                           P\sum\limits_{l\in \mathbb{Z}_{P}\setminus\mathcal{I}}\omega_{P}^{lb}, &\tau=Nb ~{\rm{with}}~P\nmid b; \\
                           0, & \hbox{otherwise.}
                         \end{array}
                       \right.
  \end{equation}
By equation (\ref{sum}), we have
$
\sum\limits_{\tau=0}^{L-1}\left|\theta_{U^{i}} (\tau)\right|^{2}=P^{3}N.
$
Therefore,
$
\theta^{2}_{a}(\mathfrak{U})\geq\frac{P^{3}N-P^{2}N^{2}}{P-1}.
$
Specially, if $\mathbb{Z}_{P}\setminus\mathcal{I}$ is a $(P,N,\lambda)$ cyclic difference set over $\mathbb{Z}_{P}$, we have $\lambda=\frac{N(N-1)}{P-1}$ and
  \begin{equation}
\theta_{U^{i}}^{2} (\tau)=P^{2}\sum\limits_{l,l'\in \mathbb{Z}_{P}\setminus\mathcal{I}}\omega_{P}^{(l-l')b}=P^{2}(N-\lambda), {\rm ~{for}}~\tau=Nb ~{\rm{with}}~P\nmid b.
  \end{equation}
The proof then follows.
 \end{IEEEproof}
\begin{remark}
Note that \textit{Construction 2} is a special case of \textit{Construction 3} when $T=1$. In particular, $\mathbb{Z}_{N+1}\setminus\{s_{0}\}$ is a $(N+1,N,N-1)$ difference set for any $s_{0}\in \mathbb{Z}_{N+1}$. Hence, we have $\theta_{a}(\mathfrak{U})=P$. It is interesting to find that the cross-correlation between different sequences is always optimal regardless of the value of $T$. Similar to \textit{Theorem 3}, the SCS family in \textit{Construction 3} is asymptotically optimal with respect to (\ref{Liubound}) when $\lim\limits_{N\rightarrow\infty}\tilde{F}(N)=\infty.$
\end{remark}

\begin{example}\label{exm22}
	Let $\mathcal{F}$ be a $4\times 5$ CFR given below:
 	\begin{equation}\label{flo5}
		\left[\begin{array}{ccccc}
			0 & 1 & 2 & 3 & 4 \\
			0 & 2 & 4 & 1 & 3 \\
			0 & 3 & 1 & 4 & 2 \\
			0 & 4 & 3 & 2 & 1
		\end{array}\right].
	\end{equation}
Here $M=4$, $N=5$, $P=N+6=11$, and $L=NP=55$. Let us choose {$\mathcal{I}=\{0,2,6,7,8,10\}$}, then according to \textit{Construction \ref{gcon}}, we get $\mathfrak{U}=\{\mathcal{U}^0,\mathcal{U}^1,\dots ,\mathcal{U}^{3}\}$, where each of the $\mathcal{U}^i$ for $0\leq i <4$, consists of a single sequence of length $55$, constructed by interleaving the base {matrix} $\mathcal{B}^i$ of order $5\times 11$, as described in our framework. A glimpse of the autocorrelation and cross-correlation of the generated sequences in time domain is shown in the first two sub-plots of Fig. \ref{fig22}. As shown in Fig. \ref{fig22}, $\theta_{a}(\mathfrak{C})=19,\theta_{c}(\mathfrak{C})=11$.
	
	As described in \textit{Construction \ref{gcon}}, the spectrum of the elements $\hat{u}^i_n$ of the frequency domain sequences $\widehat{U}^i$ for $0\leq i<4$ are
	\begin{equation}
		|\hat{u}^i_n|=\begin{cases}
			0 &\text{ for } n\in \Omega,\\
			\sqrt{\frac{11}{5}} & \text{ for } n\notin \Omega,
		\end{cases}
	\end{equation}
where $\Omega=\{  0,2,6,7,8,10,11,13,17,18,19,21,22,24,28,29,30,32,33,35,39,40,41,43,44,46,50,51,52,54\}$ is the spectral-null constraint. The third sub-plot of Fig. \ref{fig22} shows a glimpse of the magnitudes of $u^i_n$. Hence, $\sum_{i=0}^{3}|u^i_n|^2=0$ for all $n \in \Omega$. Therefore, $\mathfrak{U}$ is an SCS family with spectral-null constraint $\Omega$.

	\begin{figure}
	\includegraphics[draft=false,width=\textwidth]{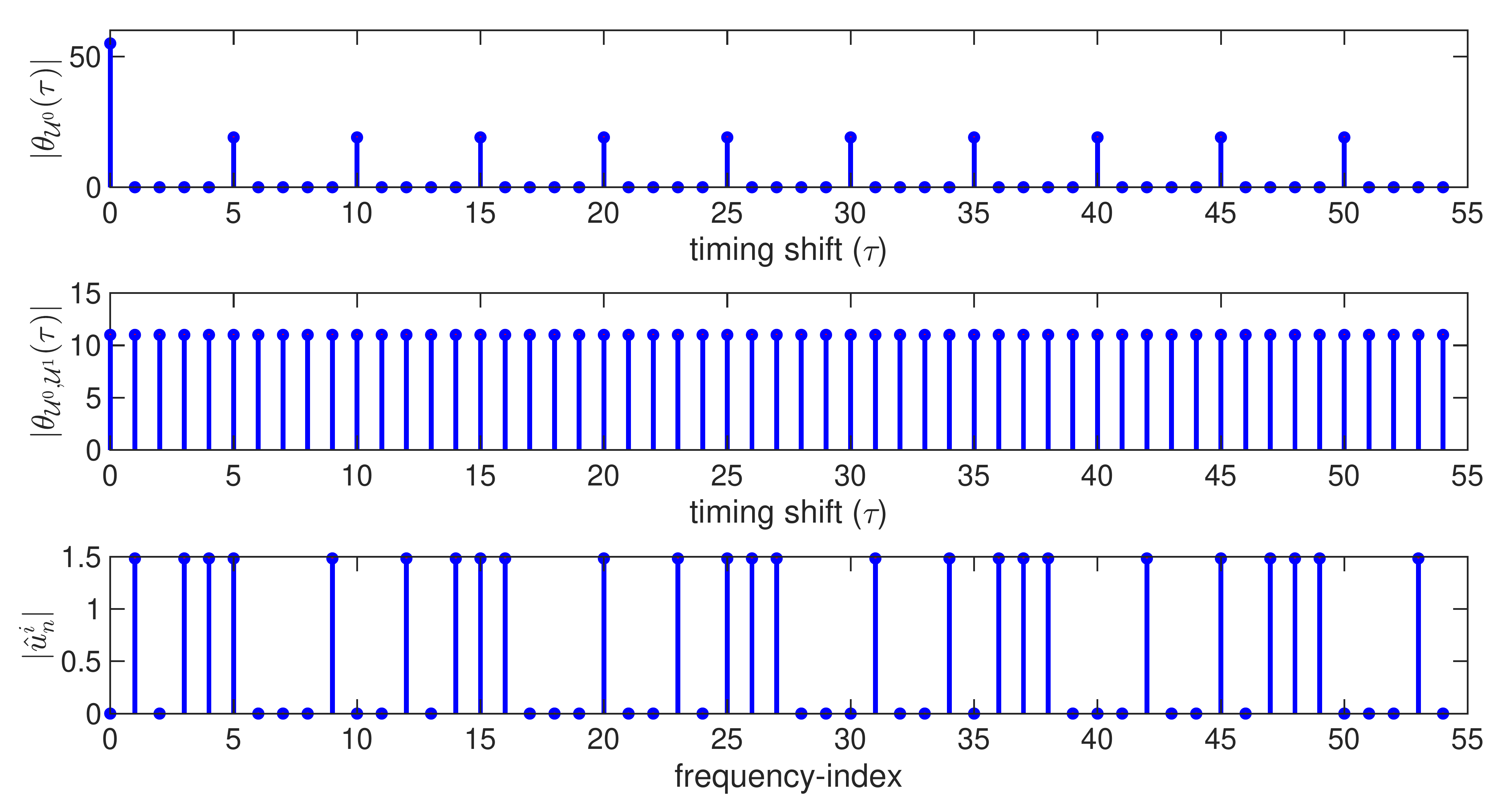}
	\caption{Illustrative plots of periodic auto- and cross- correlation magnitudes and a frequency-domain dual of the SCS family developed in \textit{Example \ref{exm22}}.}\label{fig22}
\end{figure}
\end{example}

\subsection{Optimal Multiple ZCZ SCS Sets}
In 5G physical random access cellular networks, the maximum possible delay between
different preamble sequences inside a cell is dependent on the cell range. When optimal ZCZ sequence sets are allocated to different cells, however, larger delay is possible because of the larger distance between any two diferent cells. Therefore, it is desirable to minimize the inter-set cross-correlation between different ZCZ sequence sets for minimum inter-cell interference. In this subsection, we propose a novel construction of SCS family with the aid of CFRs, which displays a large ZCZ in the time domain and minimum inter-set cross-correlation with respect to the improved lower bound in (\ref{cross-ZCZ}).

\begin{construction}\label{const3}
	Let us consider an $N\times N$ orthogonal matrix $\mathcal{H}$, such as DFT matrix, as follows
	\begin{equation}
		\mathcal{H}=\begin{bmatrix}
			h_{0,0} & h_{0,1} & \dots & h_{0,N-1}\\
			h_{1,0} & h_{1,1} & \dots & h_{1,N-1}\\
			\vdots & \vdots & \ddots & \vdots \\
			h_{N-1,0} & h_{N-1,1} & \dots & h_{N-1,N-1}
		\end{bmatrix}_{N\times N}.
	\end{equation}
	For $0\leq c <N-1$, let
	\begin{equation}
		\mathcal{H}_{c}=\begin{bmatrix}
			h_{c,0} & h_{c,1} & \dots & h_{c,N-1}\\
			h_{c,0} & h_{c,1} & \dots & h_{c,N-1}\\
			\vdots & \vdots & \ddots & \vdots \\
			h_{c,0} & h_{c,1} & \dots & h_{c,N-1}
		\end{bmatrix}_{N\times N}.
	\end{equation}
Define
	\begin{equation}
			\mathcal{A}^{i}_{c}=\mathcal{H}_{c}\circ \mathcal{A}^i,
	\end{equation}
where $\circ$ denotes the Hadamard product of matrices and {$\mathcal{A}^{i}$ is defined as equation (\ref{eq30}) and (\ref{eq29}).} By our framework, the $i$-th base matrix $\mathcal{D}^i$ of order $N \times P$ is given by $\mathcal{D}^i_{c} =\left(d^{i,c}_{j,k}\right)_{N\times P}$, where
	\begin{equation}\label{eq52}
		d^{i,c}_{j,k}=\begin{cases}
			a^i_{j,t}h_{c,t} & \text{ if }k=l_{t};\\
			0 & \text{ if }k\in \mathcal{I}.
		\end{cases}
	\end{equation}

%

Then we obtain a multiple SCS {set} consisting of $K=\tilde{F}(N)$ sets $\mathfrak{U}=\{\mathcal{U}^{i}\}_{i=0}^{K-1}$ with $\mathcal{U}^{i}=\{U^{i,c}\}_{c=0}^{N-1}$, where each $\widehat{U}^{i,c}=[\hat{u}^{i,c}_0,\hat{u}^{i,c}_1,\dots,\hat{u}^{i,c}_{NP-1}]$ is the frequency domain dual sequence with length $NP$, and
\begin{equation}\label{eq53}
	\hat{u}^{i,c}_n=d^{i,c}_{r,s} \text{ for }{r=\left\lfloor\frac{n}{P}\right\rfloor,s=n-Pr} \text{ and }0\leq n<NP.
\end{equation}
\end{construction}
For multiple SCS sets $\mathfrak{U}$ we have the following {theorem}.

\begin{theorem}\label{th4}
 $\mathfrak{U}=\{\mathcal{U}^0,\mathcal{U}^1,\dots,\mathcal{U}^{K-1}\}$ in \textit{Construction \ref{const3}} gives multiple unimodular ZCZ SCS sets which are optimal with respect to the bounds (\ref{eq10}) and (\ref{cross-ZCZ}). In summary, we have
	\begin{enumerate}
		\item The spectral constraint set of $\mathfrak{U}$ is $\Omega=\{s+aP:s\in\mathcal{I},a\in \mathbb{Z}_N\}$.
		\item The ZCZ length of $\mathcal{U}^{i}$ is $Z=N$.
        \item $\left|\theta_{U^{i,c_0},U^{i^\prime,c_1}}(\tau)\right|=P$, for any $0\leq i\neq i'\leq K-1,0\leq c_0,c_1\leq N-1$.
	\end{enumerate}
\end{theorem}	
\begin{IEEEproof}
	The first part of the theorem is similar with previous analysis, so we omit it. For $0\leq \tau<NP$, $0\leq c_0<N$ and $0\leq c_1<N$, similar to (\ref{eq35}), we have
	\begin{equation}\label{eq56}
		\begin{split}
				\theta_{U^{i,c_0},U^{i^\prime,c_1}}(\tau)&=\sum\limits_{s=0}^{N-1} \omega_{NP}^{ l_s\tau} \sum\limits_{r=0}^{N-1}a^{i}_{r,s}h_{c_0,s} {\left(a^{i^\prime}_{r,s}h_{c_1,s}\right)}^*\omega_{N}^{ r\tau}.
		\end{split}
	\end{equation}
 Now, we prove the second part of the theorem. Let us fix $i=i^\prime$, when $c_0=c_1$, then we have from (\ref{eq56}) that
\begin{equation}\label{eq57}
	|\theta_{U^{i^\prime,c_0}}(\tau)|=\begin{cases}
		NP & \text{ if }\tau=0;\\
        P\sum\limits_{s=0}^{N-1}\omega_{P}^{bl_s}, &\text{ if }\tau=Nb ~{\rm{with}}~P\nmid b; \\
        0, & \hbox{otherwise.}
	\end{cases}
\end{equation}
When $c_0\neq c_1$, we have from (\ref{eq56}) that
\begin{equation}\label{eq58}
	\theta_{U^{i^\prime,c_0},U^{i^\prime,c_1}}(\tau)=\begin{cases}
P\sum\limits_{s=0}^{N-1}h_{c_0,s}h_{c_1,s}^*  & \text{ if } \tau=0;\\
P\sum\limits_{s=0}^{N-1}h_{c_0,s}h_{c_1,s}^*\omega_{P}^{bl_s} &\text{ if }\tau=Nb ~{\rm{with}}~P\nmid b;\\
 0 & \text{ otherwise};
	\end{cases}
\end{equation}
Since $\mathcal{H}$ is an orthogonal matrix, we have $\sum\limits_{s=0}^{N-1}h_{c_0,s}h_{c_1,s}^* =0$, and hence from (\ref{eq57}) and (\ref{eq58}), we conclude that the ZCZ length of the proposed SCS $\mathcal{U}^i$ is $Z\geq N$. Since here $M=N$, and $|\Omega|=N(P-N)$, we have $Z=N$ from (\ref{eq10}) and
	\begin{equation}
		NP-|\Omega|=N^2\geq MZ.
	\end{equation}
Hence, the proposed sequence sets are optimal.

Now, we prove the third part of the theorem. From (\ref{eq56}) and (\ref{eq39}), we have

\begin{equation}\label{eq60}
		\begin{split}
				\theta_{U^{i,c_0},U^{i^\prime,c_1}}(\tau)&=\sum\limits_{s=0}^{N-1}h_{c_0,s}h^{*}_{c_1,s} \omega_{NP}^{ l_s\tau} \sum\limits_{r=0}^{N-1} \omega_{N }^{ r\left(g_{i, s}-g_{i', s}+\tau\right)}.
		\end{split}
	\end{equation}

	Based on \textit{Lemma \ref{lem5}}, $g_{i, s}-g_{i', s}+\tau\equiv0 \pmod N$ exactly has one solution $s'\not \in \mathcal{I}$ when $i\neq i^{\prime}$. If $g_{i, s}-g_{i', s}+\tau \not\equiv0 \pmod N$, then $\sum\limits_{r=0}^{N-1}\omega_{N}^{r(g_{i,s-1}-g_{i',s-1})}=0$. Hence,
\begin{equation}
	|\theta_{U^{{i},c_{0}}, U^{{i'},c_{1}}}(\tau)|=\frac{P}{N}\left[N|h_{c^\prime,s}h^{*}_{c^\prime,s}\omega_{NP}^{ l_{s^\prime}\tau}|   +0\right]=P.
\end{equation}
 In addition,  for any $0\leq i\neq i'\leq K-1,0\leq c_0,c_1\leq N-1$, we have
	\begin{equation}
\frac{NP}{\sqrt{PN-|\Omega|}}=P=\left|\theta_{U^{i,c_0},U^{i^\prime,c_1}}(\tau)\right|,
	\end{equation}
which satisfies (\ref{cross-ZCZ}) with equality, indicating that the proposed sequences families have minimum inter-set cross-correlation.
This completes the proof.
\end{IEEEproof}
%

\begin{example}\label{exm3}
	Let $\mathcal{G}$ be the $4\times 15$ CFR given in (\ref{flo15}). Here $K=4$, $N=15$, $P=N+1=16$, and $L=NP=240$. Let us choose $0\leq s_0<P(=16)$, arbitrarily, say, $s_0=4$. Consider a DFT matrix $\mathcal{H}$ of order $15$. Then, according to \textit{Construction \ref{const3}}, we get $\mathfrak{U}=\{\mathcal{U}^{0},\mathcal{U}^{1},\dots ,\mathcal{U}^{14}\}$, where $\mathcal{U}^{i}=\{U^{i,c}\}_{c=0}^{14}$ consists of a single sequence of length $240$, constructed by interleaving the base sequences $\mathcal{D}^i_c$ of order $15\times 16$, as described in (\ref{eq52}). A glimpse of the autocorrelation and cross-correlation of the generated sequences in time domain is shown in the first two sub-plots of Fig. \ref{fig3}. As shown in Fig. \ref{fig3}, the ZCZ length of $\mathcal{U}^i$ is $Z=15$.
	
	As described in (\ref{eq53}), the spectrum of the elements $\hat{u}^{i,c}_n$ of the frequency domain sequences $\widehat{U}^{i,c}$ for $0\leq c<14$ {is}
	\begin{equation}
		|\hat{u}^{i,c}_n|=\begin{cases}
			0 &\text{ for } n\in \Omega,\\
			\sqrt{\frac{16}{15}} & \text{ for } n\notin \Omega,
		\end{cases}
	\end{equation}
	where $\Omega=\{   4,~    20,~    36,~    52,~    68,~    84,~   100,~   116,~   132,~   148,~   164,~   180,~   196,~   212,~   228\}$ is the spectral-null constraint. The third sub-plot of Fig. \ref{fig3} shows a glimpse of the magnitudes of $\hat{u}^i_n$. Hence, $\sum_{c=0}^{14}|\hat{u}^{i,c}_n|^2=0$ for all $n \in \Omega$. Therefore, $\mathfrak{U}$ gives multiple SCS sets with spectral-null constraint $\Omega$.
		\begin{figure}
		\includegraphics[draft=false,width=\textwidth]{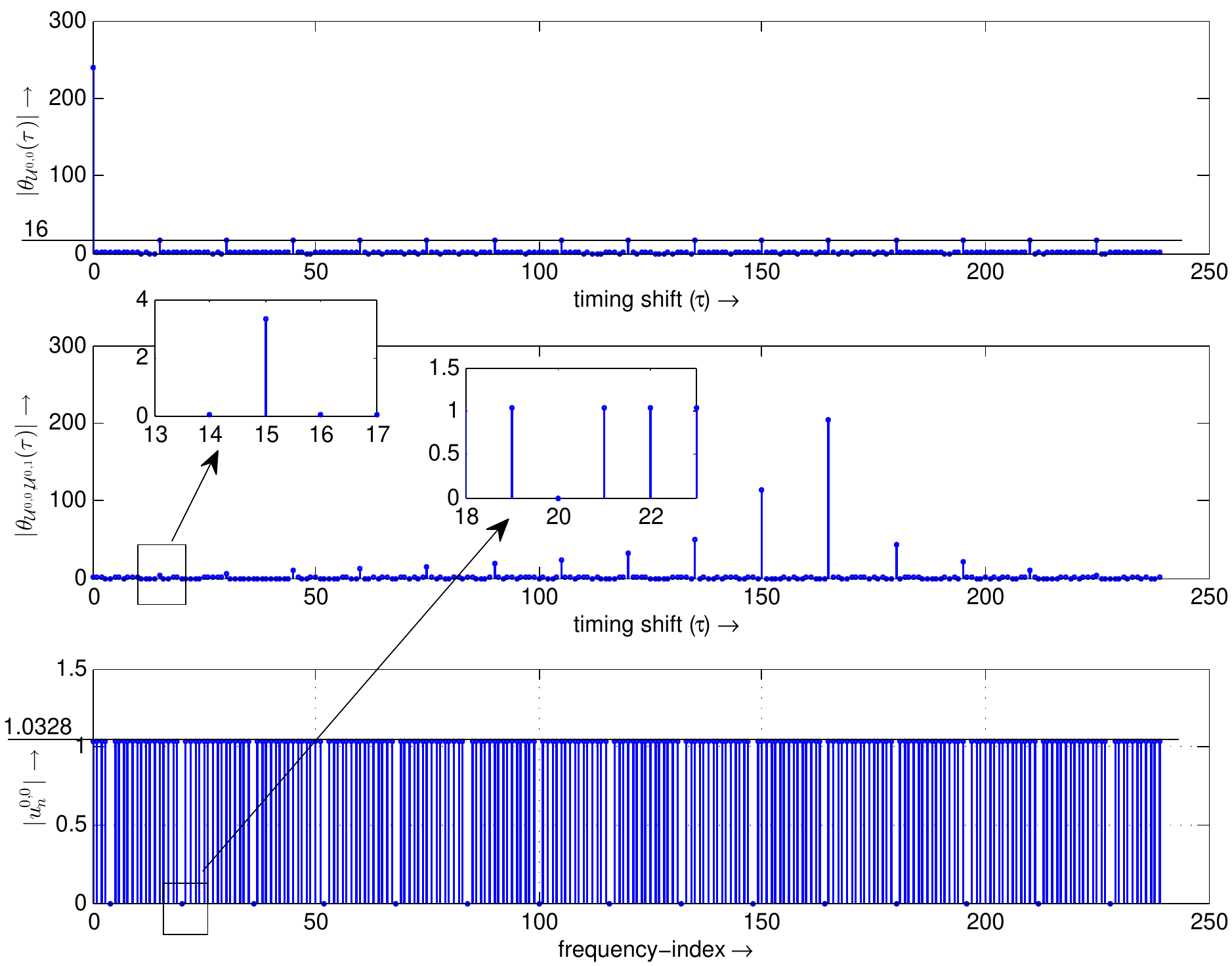}
		\caption{Illustrative plots of periodic auto- and cross- correlation magnitudes and a frequency-domain dual of the multiple SCS sets developed in \textit{Example \ref{exm3}}.}\label{fig3}
	\end{figure}
Following \textit{Theorem \ref{th4}}, we can also show that$\left|\theta_{U^{i,c_0},U^{i^\prime,c_1}}(\tau)\right|=16$, for any $0\leq i\neq i'\leq 3,0\leq c_0,c_1<15$.
\end{example}

\section{Conclusions}

In this paper, we have presented an improved periodic correlation lower bound for SCSs by differentiating the auto- and cross- correlation lower bounds of SCSs separately (see \textit{Theorem 1} in Section III). The proposed lower bound is tighter than some known bounds, such as Sarwate bounds, inequality $(44)$ in \cite{liu2018} if $M=1$ and inequality $(12)$ in \cite{tsai} for SCS set. We have also constructed a class of unimodular SCSs with uniformly low correlation sidelobes asymptotically meeting the lower bound (\ref{Liubound}) with equality (see \textit{Construction 1} in Section III). For more flexible spectral null-constraints, we have presented a unifying framework through interleaving technique in frequency domain (see Section V). Some asymptotically optimal SCS families with new spectrum holes have been proposed by selecting the base sequence based on CFRs (see \textit{Construction 2} and \textit{Construction 3} in Section VI). In particular, we have also constructed multiple SCS sets with ZCZ properties (see \textit{Construction 4} in Section VI), which are not only optimal with respect to the lower bound (\ref{eq10}), but also optimal to the newly derived bound on inter-set cross correlation in \textit{Theorem 2}. A future task of this research is to construct more optimal SCSs which have optimal correlations approaching the derived bound in \textit{Theorem 1} and \textit{Theorem 2} and explore the application scenarios of these SCSs.

\section*{Acknowledgments}

The authors are very grateful to the Associated Editor, Prof. Daniel Katz, and anonymous reviewers for their valuable comments that improved the presentation and quality of this paper.

\end{document}